\documentclass[nofootinbib, reprint,
preprintnumbers, amsmath, amssymb,
aps, prd, superscriptaddress, floatfix]{revtex4-2}
\usepackage{graphicx}

\usepackage{mathtools}
\usepackage{dcolumn}
\usepackage{bm}
\usepackage{physics}
\usepackage{hyperref}
\usepackage{braket}
\usepackage{float}

\usepackage{mathrsfs}

\usepackage{xcolor}
\usepackage{lipsum}

\begin{document}

\preprint{KEK-TH-2272}

\title{Weak Value Amplification in High Energy Physics: A Case Study for Precision Measurement of $CP$ Violation in $B$ Meson Decays}

\author{Satoshi~Higashino}
\email{higashino@people.kobe-u.ac.jp}
\affiliation{%
 Department of Physics, Kobe University, Hyogo 657-8501, Japan.
}%

\author{Yuichiro~Mori}
\email{yuichiro@post.kek.jp}
\affiliation{%
 Department of Physics, the University of Tokyo, Tokyo 113-0033, Japan.
}%

\author{Yosuke~Takubo}
\email{yosuke.takubo@kek.jp}
\affiliation{%
Institute of Particle and Nuclear Studies, High Energy Accelerator Research Organization (KEK), Ibaraki 305-0801, Japan.
}%
\affiliation{%
 The Graduate University for Advanced Studies (SOKENDAI), Hayama 240-0193, Japan.
}%

\author{Takeo~Higuchi}
\email{takeo.higuchi@ipmu.jp}
\affiliation{%
 Kavli Institute for the Physics and Mathematics of the Universe (WPI), the University of Tokyo, Kashiwa 277-8583, Japan.
}%

\author{Akimasa~Ishikawa}
\email{akimasa.ishikawa@kek.jp}
\affiliation{%
 Institute of Particle and Nuclear Studies, High Energy Accelerator Research Organization (KEK), Ibaraki 305-0801, Japan.
}%
\affiliation{%
 The Graduate University for Advanced Studies (SOKENDAI), Hayama 240-0193, Japan.
}%
\affiliation{%
International Center for Elementary Particle Physics, the University
of Tokyo, Tokyo 113-0033, Japan.
}

\author{Izumi~Tsutsui}
\email{izumi.tsutsui@kek.jp}
\affiliation{%
 Department of Physics, the University of Tokyo, Tokyo 113-0033, Japan.
}%
\affiliation{%
 Institute of Particle and Nuclear Studies, High Energy Accelerator Research Organization (KEK), Ibaraki 305-0801, Japan.
}%
\affiliation{%
 The Graduate University for Advanced Studies (SOKENDAI), Hayama 240-0193, Japan.
}%

\date{\today}

\begin{abstract}
The technique of weak value amplification, proposed by Aharonov {\it et al.}\ in 1988, has been applied for various fields of physics for the purpose of precision measurement, which is made possible by exploiting the freedom of \lq postselection\rq\ specifying actively the final state in the physical process. Here we report the feasibility of utilizing the technique of weak value amplification in high energy particle physics, especially in measuring the $CP$-violating parameters in $B$ meson decays, where the effective lifetime of the decay mode is expected to be prolonged statistically due to the postselection.  Our analysis shows that, when adopted in the Belle II experiment at the SuperKEKB collider, the effective lifetime may be prolonged up to $2.6$ times, and that the measurement precision of the $CP$-violating parameters will also be improved by its effect.
\end{abstract}


\maketitle

\section{Introduction}\label{sec:introduction}

In physical sciences, measurement is arguably the most fundamental practice to find something new in a given theoretical framework.  In quantum physics, which applies to systems of diverse scales ranging from elementary particles to condensed matter and even to the early universe, the framework is provided by Hermitian operators of observables serving as basic tools to make theoretical predictions under initially prepared quantum states.  It has been firmly believed that the basis of these predictions must be laid on the eigenvalues of the operators or their statistical average, and the measurements are designed to compare these numbers with the outcomes obtained in the actual experiments. Concomitantly, our very notion of physical reality has been tightly tied to the eigenvalues of these operators.  

An initiative to reconsider this conventional wisdom was put forward in 1988 when the novel notion of physical quantity, known as {\it weak value}, was proposed by Aharonov {\it et al.} in a time-symmetric formulation of quantum theory~\cite{Aharonov_1988}.  This formulation is an alternative but equivalent in content to the standard one, and the idea behind is quite simple: one just replaces the notion of \lq state\rq\ by \lq process\rq, that is, instead of choosing only the initial state $\ket{\psi}$ one also chooses the final state $\ket{\phi}$ and treats them equally.  On account of their active nature, the choices of the two states, $\ket{\psi}$ and $\ket{\phi}$, are referred to as {\it preselection} and {\it postselection}, respectively.
It is then expected that, given an observable $\hat{A}$, one may consider the weak value
\begin{align}
    A_{\rm w}=\frac{\braket{\phi|\hat{A}|\psi}}{\braket{ \phi|\psi}}
    \label{weakvalue_def}
\end{align}
as a tangible physical quantity, and that this weak value can be observed by the procedure called {\it weak measurement} in which one implements the postselection at the end of the measurement in addtion to the preselection made at the begining. Over the last two decades, this has been confirmed in various experiments, yielding major ramifications in two respects, one is philosophical and the other practical. 

On the philosophical side, 
the weak value offers novel interpretations of physical reality in paradoxical phenomena such as the double slit experiment involving the wave-particle duality
~\cite{Kocsis1170, Mori:2015}, the decoupling of spin and position degrees of freedom of neutrons~\cite{Denkmayr:2014aa} or the three box paradox where probabilities may become negative~\cite{Aharonov_1991} along with the related issue of the photon path~\cite{Danan_2013, Vaidman:2018}. On the practical side, which is directly related to the topic of the present paper, the weak measurement can be used for precision measurement of parameters that are otherwise difficult to detect. This intriguing possibility arises thanks to the freedom in implementing the postselection in relation to the preselection.  To be more explicit, the weak value $A_{\rm w}$ can possibly be amplified by choosing the two states $\ket{\psi}$ and $\ket{\phi}$ such that the denominator of $A_{\rm w}$ is rendered small while the numerator is kept finite.  This possibility, which was already suggested in~\cite{Aharonov_1988}, has actually been utilized in recent years for a number of purposes, including the ultrasensitive detection of beam deflection and the observation of the spin Hall effect of light~\cite{Hosten_2008, Dixon_2009}.  A more recent study reports that the sensitivity has been further improved by several orders of magnitude~\cite{Huang_2019}.   

Despite its potential universality, however, the application of weak measurement appears to have been limited mostly to optical systems so far, with the exception of neutrons~\cite{Denkmayr:2014aa} and atoms~\cite{Shomroni_2013}.  One may therefore wonder if the weak measurement can also be put to use in high energy particle physics whose energy scale is far from those previously applied.  
In fact, we notice that, even before its application in quantum optics, the weak measurement has already been in use for long time in particle physics, given that the initial and final states in particle scattering may be regarded precisely as the states chosen for the preselection and postselection.  
On the other hand, we also notice that the freedom of postselection has never been utilized for the particular purpose of amplifying the prospected signals before.  The reason for this may be ascribed to the fact that the possible range of final states in high energy experiments appears to be dictated by nature, not by experimenters, leaving no freedom of postselection to be adjusted.  
Besides, it has not been clear under what context the weak value becomes relevant in high energy physics.  

One of the aims of the present paper is to point out that in general there do exist such a context 
in systems involving decaying/oscillatory phenomena among different species of particles.  One is then 
allowed to consider the weak value of the Hamiltonian (or energy), and if the freedom of postselection can be exploited to achieve the amplification of the weak value, one finds an effective prolongation of lifetime of the decaying particle.  Obviously, such systems are not uncommon in atomic and high energy physics, with
the $B$ meson system being a prime example.  Indeed, one can recognize, in retrospect, that the weak measurement has implicitly been realized at the $B$ factories with various choices of the decay modes as final states (see, {\it e.g.}, \cite{Bevan_2014, bigi_sanda_2009, Brodzicka:2012jm}).  Another aim of this paper is to furnish a case study on the consequence of weak measurement when the freedom of postselection is fully available in the $B$ meson system used as an archetype.

More specifically, here we discuss how the lifetime of the $B_{d}$ meson can be prolonged effectively by the weak value amplification, and also argue the prospect of improving the sensitivity in measuring $CP$-violating parameters. Throughout the paper we call  $B_{d}$ meson as $B$ meson for brevity, and consider only the $CP$ violation induced by flavor mixing~\cite{PhysRevLett.45.952,PhysRevD.23.1567,Bigi:1981qs}. As the mass eigenstates of the $B$ meson ($\ket{B_{L}}$ and $\ket{B_{H}}$) are inseparable due to the magnitude of the decay width comparable with their mass difference, one can expect that the effective lifetime of the $B$ meson provides a suitable testing ground for the feasibility study of the weak measurement in particle physics. In fact, since the mass eigenstates $\ket{B_{L}}$ and $\ket{B_{H}}$ are both superpositions of the flavor eigenstates $\ket{B^{0}}$ and $\ket{\bar{B}^{0}}$,
and since the $CP$ violation manifests in the difference of decay probabilities of the initial flavor eigenstates, the effective lifetime of the mass eigenstates can be expected to be prolonged by a proper postselection of the final state $\ket{\phi}$. This may also improve the sensitivity in the measurement of $CP$-violating parameters, which should be studied together with other contributing factors that entail the postselection.

The Belle~II experiment~\cite{Abe:1304162} is operated at the SuperKEKB~\cite{Akai:2018mbz} electron-positron collider, where one of the objectives is to measure the $CP$-violating parameters in rare decays of the $B$ mesons, with the primary aim to observe new physics beyond the Standard Model (SM). In this experiment, a huge amount of $B^{0}\bar{B}^{0}$ pairs 
will be produced from the resonance decay of $\Upsilon(4S)$, which admits a big advantage in performing the required postselection of the final states under a large reduction of statistics. 

In this paper, we shall argue that the freedom of postselection may be exploited in the $B^{0} \to K^{\ast0}\gamma$ decay channel if we can choose the polarization of the photon to meet a certain consistency condition.  As a result, we find that the effective lifetime of the $B$ meson can be prolonged up to $2.6$ times, and that the precision measurement of the $CP$-violating parameters can also be improved, under the assumption of the realistic signal yield and the background contamination as well as the systematic errors used in the Belle~\cite{Brodzicka:2012jm} and Belle~II experiment~\cite{belleii_book}.

This paper is organized as follows. In Section~\ref{sec:introduction_of_Weak_Value} we first discuss how the weak value arises in the context of particle decay when the final states are chosen, and thereby see the reason why the effective lifetime of the particle can be prolonged by the weak measurement.  
We then present in Section~\ref{sec:Theoretical_model} a theoretical model of the $B$ mesons, where their effective lifetime is compared between the two cases with and without the postselection.  The realization of the choice at the decayed stage, which is a new aspect of weak measurements, is discussed for the decay channel $B^{0} \to K^{\ast0}\gamma$.  Our simulation study in the experiment at the SuperKEKB collider is given in Section~\ref{sec:expSimStudy}, where we argue that, if the present method of weak measurement is adopted in the Belle II experiment, we can expect the aforementioned significant prolongation of the effective lifetime along with the increase in the precision.
Finally, Section~\ref{sec:Conclusion} is devoted to our conclusion and discussions on remaining issues and future prospects in the line of this research.

\section{Weak Value in Particle Decay}
\label{sec:introduction_of_Weak_Value}

Before we delve into the discussion of applying the weak measurement in high energy particle physics, we describe the basic
reason why the weak value arises in analyzing the particle decay in general.  

To this end, we first note that the weak value $A_{\rm w}$ given in~\eqref{weakvalue_def} is in general a complex number, and that it may
take any value one wishes, provided that the pair of states $\ket{\psi}$ and $\ket{\phi}$ can be chosen freely.   This condition is in fact the presumption of the time-symmetric formulation~\cite{Aharonov_1988} where the process $\ket{\psi} \to \ket{\phi}$ is specified by selecting both of the states $\ket{\psi}$ and $\ket{\phi}$ independently and freely, rather than letting the latter determined from the former by causal time development.   For this reason, the states $\ket{\psi}$ and $\ket{\phi}$ are called {\it preselected state} and {\it postselected state}, respectively.  

In the context of particle decay, one may expect that the weak value amplification can be used to prolong effectively the decay time of the initial particle by 
considering the Hamiltonian operator $\hat H$ for the observable $\hat A$.  That this is indeed the case can be seen as follows.  Let $\ket{\psi}$ be the preselected state of the initial particle and $\ket{\phi}$ be the postselected state of the decayed particle(s).  The probability of transition between them during the time period $\Delta t$ will then be given by
\begin{align}
\mathrm{Pr}(\Delta t\, |\, \psi\rightarrow \phi) = \vert \langle \phi| e^{-i\Delta t\hat{H}}\ket{\psi}\vert^2.
\label{probd}
\end{align}
Suppose that the Hamiltonian $\hat{H}$ is bounded as is usually the case for considering transitions among a finite set of energy levels.  Noting that $\hat{H}$ is not Hermitian (or self-adjoint) to take account of the decaying processes, we let $m$ and $\Gamma$ be the averages of the real part and the imaginary part of the eigenvalues of $\hat{H}$, respectively. To extract only the effects of interference between the components, we may define the \lq normalized\rq\ operator of time development by 
\begin{align}
\hat{A} =\frac{2}{\Delta m}\left[\hat{H} -\left(m-\frac{i}{2}\Gamma\right)\right], 
\label{normham}
\end{align}
where ${\Delta m}$ is the range of the real part, {\it i.e.}, the difference between the largest and the smallest eigenvalues in the real part.  In particular, if there are only two states which enter in the transition, then the operator $\hat{A}$ in~\eqref{normham} may admit the form of the familiar Pauli matrices, especially $\hat{A} = \sigma_3$ in the energy diagonal basis.

Substituting $\hat{H}$ with the normalized one $\hat{A}$ in the transition amplitude, we find
\begin{align}
\langle \phi| e^{-i\Delta t\hat{H}}\ket{\psi} = e^{-\frac{1}{2}\Gamma\Delta t}e^{-im\Delta t} \langle \phi| e^{-ig\hat{A}}\ket{\psi},
\label{transamp}
\end{align}
where
\begin{align}
g = \frac{1}{2}\Delta m\Delta t =  \frac{1}{2}\left(\frac{\Delta m}{\Gamma}\right){\Gamma\Delta t}
\end{align}
is the effective coupling constant.  Now, from~\eqref{transamp} we realize that the effective range of the time period $\Delta t$ may be restricted by 
the order $O(\Gamma\Delta t) = 1$, and hence if $O(\Delta m/\Gamma) \ll 1$, then $g$ can be regarded as a small parameter $O(g) \ll 1$.  This allows us to 
employ the linear approximation,
\begin{align}
\braket{\phi| e^{-ig\hat{A}}|\psi} &\simeq \braket{\phi| I -ig\hat{A}|\psi} \nonumber\\
&= \braket{\phi|\psi} \left(1 - igA_{\rm w}\right) \simeq  \braket{\phi|\psi} e^{-igA_{\rm w}},
\label{wvrecognition}
\end{align}
where $A_{\rm w}$ is the weak value~\eqref{weakvalue_def} of the normalized operator $\hat{A}$ in~\eqref{normham}. Plugging this expression into~\eqref{probd}, we obtain 
\begin{align}
\mathrm{Pr}(\Delta t\, |\, \psi\to\phi) \simeq e^{-\Gamma\Delta t} \vert \braket{\phi|\psi} \vert^2 e^{2g\mathrm{Im}[A_{\rm w}]}.
\label{probdist}
\end{align}
This shows that, as long as the linear approximation is valid, the probability distribution of the decay may be affected by the weak value $A_{\rm w}$ (its imaginary part). Notably, by choosing the combination of the preselection and the postselection properly, the weak value $A_{\rm w}$ can be altered freely; for instance, $\mathrm{Im}[A_{\rm w}]$ may be rendered large by choosing $\braket{\phi|\psi}$ small as stated in Section~\ref{sec:introduction}.   Even if the coupling constant $g$ cannot be regarded small, the system may allow one to evaluate the transition amplitude in full order to see if the selections affect the outcome.  Later, we shall see explicitly that the decay of the $B$ meson falls into that particular case.

\section{Theoretical basis}\label{sec:Theoretical_model}
In this section, we present our theoretical basis for applying the weak measurement for the detection of $CP$ violation in the $B$ meson system.  
In our case, the Hamiltonian $\hat H$ takes the role of the observable $\hat A$, and we shall see that the imaginary part of the weak value $A_{\rm w}$ is related to the effective lifetime of the $B$ meson.  The aim of the weak measurement is then to find a source of $CP$ violation in the conditional time distribution when our state is $\ket{B^{0}}$ at the initial time and is an arbitrary chosen state $\ket{B_{\rm decay}}$ at the time of the decay (see~\eqref{defbf}). 

\subsection{Time evolution}
The state $|\psi\rangle$ which describes a neutral $B$ meson is in general a superposition of  the flavor eigenstates $|B^{0}\rangle$ and $|\bar{B}^{0}\rangle$ which are the $CP$ conjugate to each other, that is, $|\bar{B}^{0}\rangle = \mathcal{CP} |B^{0}\rangle$ and $|B^{0}\rangle = \mathcal{CP} |\bar{B}^{0}\rangle$  under the $CP$ transformation $\mathcal{CP}$ for which $(\mathcal{CP})^2 = I$. 
Since mesons are unstable particles, the time evolution is phenomenologically described by a non-Hermitian Hamiltonian $\hat{H}$. The eigenstates of the Hamiltonian are given by
\begin{align}
\hat{H}|B_{L}\rangle&=\left(m_{L}-\frac{i}{2}\Gamma_{L}\right)|B_{L}\rangle,
\label{eigenvector}\\
\hat{H}|B_{H}\rangle&=\left(m_{H}-\frac{i}{2}\Gamma_{H}\right)|B_{H}\rangle,
\label{eigenvector2}
\end{align}
where $m_{L},m_{H},\Gamma_{L},\Gamma_{H}$ are real positive numbers satisfying
\begin{align}
    m_{L}<m_{H}.
\end{align}

These eigenstates $|B_{L}\rangle$ and $|B_{H}\rangle$ can be written as 
\begin{align}
\ket{B_{L}}&=p|B^{0}\rangle+q|\bar{B}^{0}\rangle,
\label{blfromb0b0bar}\\
\ket{B_{H}}&=p|B^{0}\rangle-q|\bar{B}^{0}\rangle,
\label{bhfromb0b0bar}
\end{align}
where $p, q$ are complex numbers fulfilling $|p|^{2}+|q|^{2}=1$. It should be noted that these eigenstates are not orthogonal, and hence the Hamiltonian $\hat{H}$ cannot be diagonalized.

Using~\eqref{blfromb0b0bar} and~\eqref{bhfromb0b0bar}, we find
\begin{align}
|B^{0}\rangle&=\frac{1}{2p}\left(\ket{B_{L}}+\ket{B_{H}}\right),
\label{b0fromblbh}\\
|\bar{B}^{0}\rangle&=\frac{1}{2q}\left(\ket{B_{L}}-\ket{B_{H}}\right).
\label{b0barfromblbh}
\end{align}
In high energy experiments, we decide whether the initial state is the flavor eigenstate $\ket{B^{0}}$ or its $CP$ conjugate $\ket{\bar{B}^{0}}$ through a process called the flavor tagging.
In this paper, the time at which this tagging is performed is referred to as \lq initial\rq, and the time lapse from the initial time will be denoted by $\Delta t$.  
When the meson is identified as $B^{0}$ at the initial time, it evolves into the state
\begin{align}
|B^{0}(\Delta t)\rangle&=e^{-i\Delta t\hat{H}}|B^{0}\rangle
\label{b0-state}
\end{align}
at the time of the decay. Similarly, when the particle is identified as $\bar{B}^{0}$, we have
\begin{align}
|\bar{B}^{0}(\Delta t)\rangle&=e^{-i\Delta t\hat{H}}|\bar{B}^{0}\rangle.
\label{b0bar-state}
\end{align}
In our discussion, for the preselected state $\ket{\psi}$ we primarily choose $\ket{\psi}=\ket{B^{0}}$ except when $\ket{\bar{B}^{0}}$ is considered for comparison. Our choice for the postselected state $\ket{\phi}$ will be given later in Subsection C.

For our later convenience, we note here that, as long as the dynamical development is concerned, the evolution~\eqref{b0bar-state}  of the $CP$ conjugate $\bar{B}^{0}$ is obtained immediately from~\eqref{b0-state} by interchanging $p \leftrightarrow q$.  This can be readily confirmed from~\eqref{eigenvector} and~\eqref{eigenvector2} by expressing the Hamiltonian in terms of the parameters $p$, $q$ in the orthogonal basis of the flavor eigenstates $|B^{0}\rangle$ and $|\bar{B}^{0}\rangle$ as
\begin{align}
\hat H(p, q) &= (m - \frac{i}{2} \Gamma)\left(|B^{0}\rangle \langle B^{0}|+|\bar{B}^{0}\rangle \langle\bar{B}^{0}|\right) \nonumber \\
&\!\!\!\!\!\!\!\!\!- (\Delta m - \frac{i}{2} \Delta\Gamma)\left(\frac{p}{2q}|B^{0}\rangle \langle \bar{B}^{0}|+\frac{q}{2p}|\bar{B}^{0}\rangle \langle {B}^{0}|\right),\label{hami-b-meson}
\end{align}
using
\begin{align}
m=\frac{m_{L}+m_{H}}{2},\ \quad \Gamma=\frac{\Gamma_{L}+\Gamma_{H}}{2},
\label{avpar}
\end{align}
and 
\begin{align}
\Delta m=m_{H}-m_{L},\ \quad \Delta \Gamma=\Gamma_{H}-\Gamma_{L}.
\label{avpartwo}
\end{align}
One then immediatety observes that the $CP$ conjugation 
of the Hamiltonian results in the interchange of the parameters, $\mathcal{CP}\hat H(p, q) \,\mathcal{CP} = \hat H(q, p)$.  It then
follows that, given the transition amplitude $T(p, q) = \langle \alpha| e^{-i\Delta t\hat{H}(p, q)}|\beta\rangle$, the corresponding amplitude of the $CP$ conjugate has the same property,
\begin{align}
\bar{T}(p, q) &= \langle \bar \alpha| e^{-i\Delta t\hat{H}(p, q)}|\bar \beta\rangle 
=  \langle  \alpha| \mathcal{CP} e^{-i\Delta t\hat{H}(p, q)}\mathcal{CP} |\beta\rangle \nonumber \\
&=  \langle  \alpha|  e^{-i\Delta t\,\mathcal{CP}\hat{H}(p, q)\,\mathcal{CP}} |\beta\rangle =T(q, p),
\label{pq-exchange}
\end{align}
as claimed.

\subsection{Time distribution without postselection}
Due to the instability of the particle, the norm of the particle's state is not always unity, as can be seen explicitly as 
\begin{widetext}
\begin{align}
\langle B^{0}(\Delta t)|B^{0}(\Delta t)\rangle&=\langle B^{0}|e^{i\Delta t\hat{H}^{\dag}}e^{-i\Delta t\hat{H}}|B^{0}\rangle \nonumber\\
&=\frac{1}{4|p|^{2}}\left(e^{im_{L}\Delta t-\frac{\Gamma_{L}}{2}\Delta t}\bra{B_{L}}+e^{im_{H}\Delta t-\frac{\Gamma_{H}}{2}\Delta t}\bra{B_{H}}\right)
\cdot\left(e^{-im_{L}\Delta t-\frac{\Gamma_{L}}{2}\Delta t}\ket{B_{L}}+e^{-im_{H}\Delta t-\frac{\Gamma_{H}}{2}\Delta t}\ket{B_{H}}\right) \nonumber\\
&=\frac{1}{4|p|^{2}}\left(e^{-\Gamma_{L}\Delta t}+e^{-\Gamma_{H}\Delta t}\right)\quad+\frac{e^{-\Gamma\Delta t}}{4|p|^{2}}\left(e^{i\Delta m\Delta t}\langle B_{L}|B_{H}\rangle + e^{-i\Delta m\Delta t}\langle B_{H}|B_{L}\rangle\right) \nonumber\\
&=\frac{1}{4|p|^{2}}\left(e^{-\Gamma_{L}\Delta t}+e^{-\Gamma_{H}\Delta t}\right)+\frac{|p|^{2}-|q|^{2}}{2|p|^{2}}e^{-\Gamma\Delta t}\cos\left(\Delta m\Delta t\right). 
\label{decay_rat}
\end{align}
 Similarly, when the initial state is $|\bar{B}^{0}\rangle$, we find
\begin{align}
    \langle \bar{B}^{0}(\Delta t)|\bar{B}^{0}(\Delta t)\rangle&=\frac{1}{4|q|^{2}}\left(e^{-\Gamma_{L}\Delta t}+e^{-\Gamma_{H}\Delta t}\right)-\frac{|p|^{2}-|q|^{2}}{2|q|^{2}}e^{-\Gamma\Delta t}\cos\left(\Delta m\Delta t\right).
    \label{decay_rat_b0bar}
\end{align}
\end{widetext}
To clarify the meaning of~\eqref{decay_rat}, we insert the identity
\begin{align}
I=|B^{0}\rangle\bra{B^{0}}+|\bar{B}^{0}\rangle\langle\bar{B}^{0}|
\end{align}
into the left hand side of~\eqref{decay_rat} to obtain
\begin{align}
\langle B^{0}(\Delta t)|B^{0}(\Delta t)\rangle&=|\langle B^{0}|B^{0}(\Delta t)\rangle|^{2}+|\langle \bar{B}^{0}|B^{0}(\Delta t)\rangle|^{2}.
\label{meaning_of_amp}
\end{align}
This shows that~\eqref{decay_rat} provides the probability of remaining either as $B^{0}$ or $\bar{B}^{0}$ without decaying until the time $\Delta t$ after the particle starts as $B^{0}$ at the initial time.  Note that the result~\eqref{decay_rat_b0bar} is also obtained from~\eqref{decay_rat} by the interchange $p \leftrightarrow q$ as we noted before.  

Taking the normalization condition of probability into account, we observe that the time distribution $P(\Delta t)$ expressing the probability density of decay at time $\Delta t$ reads
\begin{align}
P(\Delta t)
&=-\left.\frac{d}{d\Delta t'}\langle B^{0}(\Delta t')|B^{0}(\Delta t')\rangle\right|_{\Delta t' =\Delta t} \nonumber \\
&=\frac{1}{4|p|^{2}}\left(\Gamma_{L}e^{-\Gamma_{L}\Delta t}+\Gamma_{H}e^{-\Gamma_{H}\Delta t}\right)\nonumber\\
&\quad +\frac{|p|^{2}-|q|^{2}}{2|p|^{2}}e^{-\Gamma\Delta t}\Delta m \sin\left(\Delta m\Delta t\right)\nonumber\\
&\quad +\frac{|p|^{2}-|q|^{2}}{2|p|^{2}}e^{-\Gamma\Delta t}\Gamma\cos\left(\Delta m\Delta t\right).  
\label{default_time_dis}
\end{align}
We note in passing that, if we implement the postselection further, we will obtain the conditional probability $\mathrm{Pr}(\Delta t|\psi\to\phi)$ in~\eqref{probd}. The result $P(\Delta t)$ in~\eqref{default_time_dis} will then be derived, up to normalization, by summing it up for all $\ket{\phi}$ states.

The effective lifetime of this particle can then be evaluated as
\begin{align}
\tau_{\rm eff}(B^{0})&=\int_{0}^{\infty} d\Delta t'\ \Delta t' P(\Delta t')\nonumber\\
&=-\int_{0}^{\infty}d\Delta t'\  \Delta t'\left[\frac{d}{d \tau }\langle B^{0}(\tau)|B^{0}(\tau)\rangle\right]_{\tau =\Delta t'}  \nonumber\\
&=+\left[\Delta t' \langle B^{0}(\Delta t')|B^{0}(\Delta t')\rangle\right]^{\Delta t'=\infty}_{\Delta t'=0}  \nonumber\\
&\quad+\int_{0}^{\infty}d\Delta t'\ \langle B^{0}(\Delta t')|B^{0}(\Delta t')\rangle  \nonumber\\
&=\frac{1}{4|p|^{2}}\left(\frac{1}{\Gamma_{L}}+\frac{1}{\Gamma_{H}}\right)+\frac{|p|^{2}-|q|^{2}}{2|p|^{2}}\frac{\Gamma}{\Gamma^{2}+\Delta m^{2}}. 
\label{lifetime_without_postselection}
\end{align}
Since the actual values of $\Gamma_{L}$ and $\Gamma_{H}$ are both close to $\Gamma$, it is assuring to observe that, for $|p|\approx|q|\approx 1/\sqrt{2}$, we have $\tau_{\rm eff}(B^{0})\approx 1/\Gamma$, which is the lifetime of the $B^{0}$ meson.

\subsection{Time distribution with postselection}

Now we wish to implement the postselection and see how it affects the time distribution obtained before without the postselection.  
The postselected state $\ket{\phi}$ can in general be written as a superposition of the two states, $|B^{0}\rangle$ and $|\bar{B}^{0}\rangle$. As we argue later in Subsection E, such a postselected state may be realized as a state $\ket{B_{\rm decay}}$ from a particular decay mode $f$ measured in the experiment. This will be our postselected state, namely, our postselected state $\ket{\phi}=\ket{B_{\rm decay}}$ is given by
\begin{align}
\ket{B_{\rm decay}}&=r|B^{0}\rangle+s|\bar{B}^{0}\rangle,
\label{defbf}
\end{align}
where $r, s$ are complex numbers satisfying the normalization condition,
\begin{align}
    |r|^{2}+|s|^{2}=1.
    \label{norm_cond_ps}
\end{align}
These two parameters $r$ and $s$ are assumed to be chosen freely by the measurement process of the decay mode $f$. In addition, when investigating $CP$ violation, the $CP$ conjugate of the state in~\eqref{defbf},
\begin{align}
    |\bar{B}_{\rm decay}\rangle=s|B^{0}\rangle+ r|\bar{B}^{0}\rangle,
    \label{defbfbar}
\end{align}
should also be considered. From now on, $\Delta t$ will be considered to be the time of the decay, which implies that $\ket{B^{0}(\Delta t)}$ and $\ket{\bar{B}^{0}(\Delta t)}$ in~\eqref{b0-state} and~\eqref{b0bar-state} are now the states realized by dynamical evolution at the time of the decay.

Before evaluating the time distribution with postselection, we first obtain the probability of finding the postselected state $|B_{\rm decay}\rangle$ specified in~\eqref{defbf} when
the state under consideration is the $B$ meson state $|B^{0}(\Delta t)\rangle$ given in~\eqref{b0-state}. The result is
\begin{widetext}
\begin{align}
\left|\langle B_{\rm decay}|B^{0}(\Delta t)\rangle\right|^{2}&=\left|r^{\ast}\langle B^{0}|B^{0}(\Delta t)\rangle+s^{\ast}\langle \bar{B}^{0}|B^{0}(\Delta t)\rangle\right|^{2} \nonumber \\
&=\left|\left(\frac{r^{\ast}}{2}+\frac{qs^{\ast}}{2p}\right)e^{-\frac{\Gamma_{L}}{2}\Delta t}e^{-i\Delta tm_{L}}+\left(\frac{r^{\ast}}{2}-\frac{qs^{\ast}}{2p}\right)e^{-\frac{\Gamma_{H}}{2}\Delta t}e^{-i\Delta tm_{H}}\right|^{2}.
\label{prob_bdecay}
\end{align}
\end{widetext}
Introducing the relative phase parameters $\varphi$ and $\theta$ through 
\begin{align}
\frac{p}{q} = \frac{|p|}{|q|} e^{i\varphi}, \qquad \frac{r}{s} = \frac{|r|}{|s|} e^{i\theta}, 
\label{pqvarphi}
\end{align}
we find
\begin{align}
&\left|\langle B_{\rm decay}|B^{0}(\Delta t)\rangle\right|^{2}\nonumber\\
&=\left(\frac{|r|^2}{2}+\frac{|q|^{2}|s|^{2}}{2|p|^2}\right)\frac{e^{-\Gamma_{L}\Delta t}+e^{-\Gamma_{H}\Delta t}}{2} \nonumber\\
&\quad+\left(\frac{|r|^{2}}{2}-\frac{|q|^2|s|^2}{2|p|^{2}}\right)e^{-\Gamma\Delta t}\cos\left(\Delta m\Delta t\right) \nonumber\\
&\quad+\frac{|r||q||s|}{|p|}\cos\left(\theta-\varphi\right)\frac{e^{-\Gamma_{L}\Delta t}-e^{-\Gamma_{H}\Delta t}}{2} \nonumber\\
&\quad-\frac{|r||q||s|}{|p|}\sin\left(\theta-\varphi\right)e^{-\Gamma\Delta t}\sin\left(\Delta m\Delta t\right).
\label{bfb0t2}
\end{align}
The corresponding $CP$ conjugate $\left|\langle \bar{B}_{\rm decay}|\bar{B}^{0}(\Delta t)\rangle\right|^{2}$ is then obtained simply by putting $|p|\leftrightarrow |q|$ and $\varphi\leftrightarrow -\varphi$ in~\eqref{bfb0t2}.

It is notable that either the second or the fourth term in~\eqref{bfb0t2} is nonvanishing for any combinations of the two parameters $(r,s)$ specifying the postselection except for the case $|p|=|q|$. These terms have oscillatory property which can be controlled by the parameters $|r|$ and $\theta$.

The time distribution with postselection is then defined by
\begin{align}
P(\Delta t| B^{0}\rightarrow B_{\rm decay})
=\frac{\left|\langle B_{\rm decay}|B^{0}(\Delta t)\rangle\right|^{2}}{\int_{0}^{\infty} d\Delta t' \left|\langle B_{\rm decay}|B^{0}(\Delta t' )\rangle\right|^{2} }
\label{CTime_distribution}
\end{align}
which is the conditional time distribution with $B^{0}$ at the initial time and $B_{\rm decay}$ at the time of the decay. 
Similarly, its $CP$ conjugate ($\bar{B}^{0}\rightarrow \bar{B}_{\rm decay}$) version is also defined by
\begin{align}
P(\Delta t| \bar{B}^{0}\rightarrow \bar{B}_{\rm decay})
=\frac{\left|\langle \bar{B}_{\rm decay}|\bar{B}^{0}(\Delta t)\rangle\right|^{2}}{\int_{0}^{\infty} d\Delta t' \left|\langle \bar{B}_{\rm decay}|\bar{B}^{0}(\Delta t')\rangle\right|^{2}}.
\label{CTime_distribution_bar}
\end{align}

Using~\eqref{bfb0t2}, we obtain the denominator of~\eqref{CTime_distribution} as
\begin{align}
&\int_{0}^{\infty} d\Delta t' \left|\langle B_{\rm decay}|B^{0}( \Delta t' )\rangle\right|^{2}\nonumber\\
&=\left(\frac{|r|^2}{2}+\frac{|q|^{2}|s|^{2}}{2|p|^2}\right)\left(\frac{1}{2\Gamma_{L}}+\frac{1}{2\Gamma_{H}}\right)\nonumber\\
&\quad+\left(\frac{|r|^{2}}{2}-\frac{|q|^2|s|^2}{2|p|^{2}}\right)
\frac{\Gamma}{\Gamma^{2}+\Delta m^{2}}\nonumber\\
&\quad+\frac{|r||q||s|}{|p|}
\cos\left(\theta-\varphi\right)\left(\frac{1}{2\Gamma_{L}}-\frac{1}{2\Gamma_{H}}\right)\nonumber\\
&\quad-\frac{|r||q||s|}{|p|}
\sin\left(\theta-\varphi\right)\frac{\Delta m}{\Gamma^{2}+\Delta m^{2}}.
\label{normalize_factor}
\end{align}

In particular, when the decay widths of the two mass eigenstates of the meson are considered to be equivalent,  $\vert\Gamma_{L}-\Gamma_{H}\vert/\Gamma \ll 1$ (we have $\vert\Gamma_{L}-\Gamma_{H}\vert/\Gamma < 0.01$ according to~\cite{PDG_2020}), the time distribution~\eqref{CTime_distribution} admits the form,
\begin{widetext}
\begin{align}
    P(\Delta t| B^{0}\rightarrow B_{\rm decay})
    &=\Gamma e^{-\Gamma\Delta t}\frac{\left(\frac{|r|^2}{2}+\frac{|q|^{2}(1-|r|^{2})}{2|p|^2}\right)
+\left(\frac{|r|^2}{2}-\frac{|q|^2(1-|r|^2)}{2|p|^{2}}\right)\cos\left(\Delta m\Delta t\right)-\frac{|q||r|\sqrt{1-|r|^2}}{|p|}\sin\left(\theta-\varphi\right)\sin\left(\Delta m\Delta t\right)}
{\left(\frac{|r|^2}{2}+\frac{|q|^{2}(1-|r|^{2})}{2|p|^2}\right)
+\left(\frac{|r|^2}{2}-\frac{|q|^2(1-|r|^2)}{2|p|^{2}}\right)
\frac{\Gamma^{2}}{\Gamma^{2}+\Delta m^{2}}
-\frac{|q||r|\sqrt{1-|r|^2}}{|p|}
\sin\left(\theta-\varphi\right)\frac{\Gamma\Delta m}{\Gamma^{2}+\Delta m^{2}}}.
\label{time-dist-wm}
\end{align}
\end{widetext}

\begin{figure}[t]
    \centering
    \includegraphics[width=8.5cm]{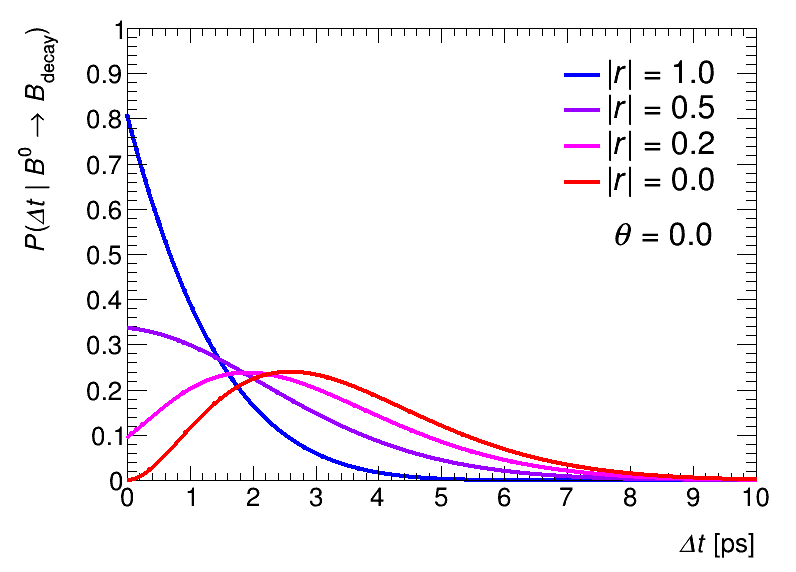}
    \caption{The time distribution $P(\Delta t|B^{0}~\to~B_{\rm decay})$ for various $|r|$ in the case of $|p| = |q|$, $\Delta m = 0.506$~$(\mathrm{ps})^{-1}$, $1 / \Gamma = 1.519$~ps, and $\theta = 0$. Each distribution is normalized to unity.}
    \label{fig:pdf_B_decay}
\end{figure}

Figure \ref{fig:pdf_B_decay} shows $P(\Delta t|B^{0}~\to~B_{\rm decay})$ for various $|r|$ values in the case of $|p|~=~|q|$, $\Delta m~=~0.506$ $(\mathrm{ps})^{-1}$, $1 / \Gamma~=~1.519 $ ps, and the postselection parameter $\theta$ is fixed to be 0. The time distribution is clearly shifted to right as $|r|$ goes to zero, indicating the increase of the effective lifetime.

\subsection{Weak value amplification}

Having seen that different time distributions can be obtained depending on whether the postselection was performed or not, we now evaluate the effective lifetime of the $B$ meson with the postselection. 
Actually, the mean value of the time distribution, which is the effective lifetime with the condition $B^{0}\rightarrow B_{\rm decay}$, is obtained by
\begin{align}
\tau_{\rm eff}(B^{0}\rightarrow B_{\rm decay})&=\int_{0}^{\infty} d\Delta t' \  \Delta t' P(\Delta t' | B^{0}\rightarrow B_{\rm decay}),\nonumber \\
&=\frac{\int_{0}^{\infty} d\Delta t' \ \Delta t'\left|\langle B_{\rm decay}|B^{0}(\Delta t' )\rangle\right|^{2}}{\int_{0}^{\infty} d\Delta t' \left|\langle B_{\rm decay}|B^{0}(\Delta t')\rangle\right|^{2} }.
\label{cond_lif_tim}
\end{align}
The denominator of~\eqref{cond_lif_tim} is given by~\eqref{normalize_factor}. The numerator is evaluated as
\begin{align}
&\int_{0}^{\infty} d\Delta t' \ \Delta t' \left|\langle B_{\rm decay}|B^{0}(\Delta t' )\rangle\right|^{2}\nonumber\\
&=\left(\frac{|r|^2}{2}+\frac{|q|^{2}|s|^{2}}{2|p|^2}\right)\left(\frac{1}{2\Gamma_{L}^{2}}+\frac{1}{2\Gamma_{H}^{2}}\right)\nonumber\\
&\quad+
\left(\frac{|r|^{2}}{2}-\frac{|q|^{2}|s|^{2}}{2|p|^{2}}\right)
\frac{\Gamma^{2}-\Delta m^{2}}{\left(\Gamma^{2}+\Delta m^{2}\right)^{2}}\nonumber\\
&\quad+
\frac{|r||q||s|}{|p|}
\cos\left(\theta-\varphi\right)\left(\frac{1}{2\Gamma_{L}^{2}}-\frac{1}{2\Gamma_{H}^{2}}\right)\nonumber\\
&\quad-
\frac{|r||q||s|}{|p|}
\sin\left(\theta-\varphi\right)\frac{2\Gamma\Delta m}{\left(\Gamma^{2}+\Delta m^{2}\right)^{2}}.\label{lifet_pre}
\end{align}

Since $\vert\Gamma_{L}-\Gamma_{H}\vert/\Gamma \ll 1$, we are allowed to put $\Gamma_{L}=\Gamma_{H}=\Gamma$ approximately. The effective lifetime~\eqref{cond_lif_tim} then admits the closed form,
\begin{widetext}
\begin{align}
\tau_{\rm eff}(B^{0}\rightarrow B_{\rm decay})&=\frac{\left(1+\frac{|q|^{2}|s|^{2}}{|p|^{2}|r|^{2}}\right)\frac{1}{\Gamma^{2}}+
\left(1-\frac{|q|^{2}|s|^{2}}{|p|^{2}|r|^{2}}\right)
\frac{\Gamma^{2}-\Delta m^{2}}{\left(\Gamma^{2}+\Delta m^{2}\right)^{2}}
+\frac{|q||s|}{|p||r|}
\sin\left(\theta-\varphi\right)\frac{4\Gamma\Delta m}{\left(\Gamma^{2}+\Delta m^{2}\right)^{2}}}
{\left(1+\frac{|q|^{2}|s|^{2}}{|p|^{2}|r|^{2}}\right)\frac{1}{\Gamma}
+\left(1-\frac{|q|^2|s|^2}{|p|^{2}|r|^{2}}\right)
\frac{\Gamma}{\Gamma^{2}+\Delta m^{2}}
+\frac{|q||s|}{|p||r|}
\sin\left(\theta-\varphi\right)\frac{2\Delta m}{\Gamma^{2}+\Delta m^{2}}},\nonumber\\
&=\frac{\left(1+\left|A_{\rm w}\right|^{2}\right)\frac{1}{\Gamma^{2}}+\left(1-\left|A_{\rm w}\right|^{2}\right)\frac{\Gamma^{2}-\Delta m^{2}}{\left(\Gamma^{2}+\Delta m^{2}\right)^{2}}+4\mathrm{Im}\left[A_{\rm w}\right]\frac{\Gamma\Delta m}{\left(\Gamma^{2}+\Delta m^{2}\right)^{2}}}{\left(1+\left|A_{\rm w}\right|^{2}\right)\frac{1}{\Gamma}+\left(1-\left|A_{\rm w}\right|^{2}\right)\frac{\Gamma}{\Gamma^{2}+\Delta m^{2}}+2\mathrm{Im}\left[A_{\rm w}\right]\frac{\Delta m}{\Gamma^{2}+\Delta m^{2}}}.
\label{life_time_weak_value}
\end{align}
\end{widetext}
Here, $A_{\rm w}$ is the weak value given in~\eqref{weakvalue_def}, which now takes the form,
\begin{align}
A_{\rm w} = \frac{|q||s|}{|p||r|}\cos(\theta-\varphi)+i\frac{|q||s|}{|p||r|}\sin(\theta-\varphi),
\end{align}
upon using $\ket{\psi}=\ket{B^{0}}$, $\ket{\phi}=\ket{B_{\rm decay}}$ and $\hat{A}$ defined in~\eqref{normham} together with~\eqref{hami-b-meson}.

\begin{figure}[h]
    \centering
    \includegraphics[width=8.5cm]{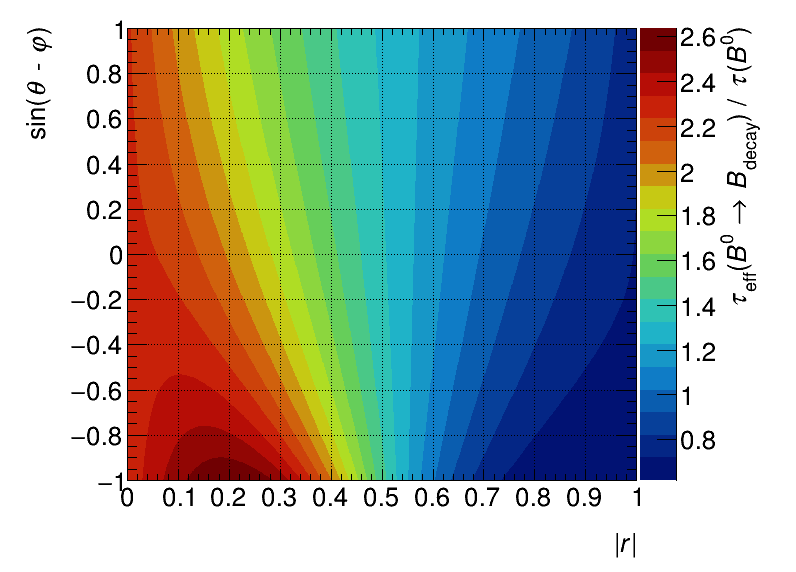}
    \caption{The ratio between the lifetime of $B$ meson $\tau(B^0)$ and the effective lifetime~\eqref{life_time_weak_value}. The effective lifetime can be extended about $2.6$ times larger than $\tau(B^{0})$ in this case when $|r|\sim 0.2$ and $\sin(\theta-\varphi)\sim -1$.}
    \label{fig:weak_value_amplification_of_lifetime}
\end{figure}

Note that we have obtained the result~\eqref{life_time_weak_value} without resorting to the first-order approximation for $\Delta m/\Gamma$. In fact, the full calculation is required for us here since $\Delta m/\Gamma \simeq 0.77$ for the $B$ meson system we are considering~\cite{PDG_2020}. 
For $B$ mesons, Figure \ref{fig:weak_value_amplification_of_lifetime} shows that this effective lifetime can be $2.6$ times larger than the lifetime $\tau(B^{0})$.

Returning to the general case, we note that when $\Delta m/\Gamma \ll 1$ holds, one may expand the effective lifetime~\eqref{life_time_weak_value} in terms of $\Delta m/\Gamma$ to obtain
\begin{align}
\tau_{\rm eff}(B^{0}\rightarrow B_{\rm decay}) =  \frac{1}{\Gamma}+ \frac{1}{\Gamma}\,\mathrm{Im}[A_{\rm w}]\frac{\Delta m}{\Gamma}+O\left[(\Delta m/\Gamma)^{2}\right].
    \label{forder_lifetime}
\end{align}
The extension of the effective lifetime is proportional to the imaginary part of the weak value $A_{\rm w}$ in the first order of $\Delta m/\Gamma$, which can also be confirmed directly from~\eqref{probdist}.  

This amplification effect can also be seen in the detection of $CP$ violation. For simplicity, we assume $|p|=|q|$ and $\Delta m/\Gamma \ll 1$ so that the difference between the denominator of~\eqref{time-dist-wm} and its counterpart for the $CP$ conjugate is small enough to be ignored. Then, the difference between their time distributions is found as
\begin{align}
   &P(\Delta t| B^{0}\rightarrow B_{\rm decay})-P(\Delta t| \bar{B}^{0}\rightarrow \bar{B}_{\rm decay})\nonumber\\
   &\propto -\frac{\sqrt{1-|r|^{2}}}{|r|}\left[\sin(\theta-\varphi)-\sin(\theta+\varphi)\right]e^{-\Gamma\Delta t}\sin(\Delta m\Delta t)\nonumber\\
   &=2\frac{\sqrt{1-|r|^{2}}}{|r|}\cos\theta\sin\varphi e^{-\Gamma\Delta t}\sin(\Delta m\Delta t).
\end{align}
This result indicates that the parameters $|r|$ and $\theta$ introduced in our postselection can be used to amplify the difference in the conditional time distribution.

\subsection{Process of postselection and related issues}
\label{subsec:proc_of_pos}
In order to perform the above amplification, we need to give an operation to choose the state $\ket{B_{\rm decay}}$. This will be performed by identifying the type of particles produced at the $B$ meson decay.

To be explicit, let $\ket{f}$ be the state of decayed particles which we identify from the measurement. The state of the $B$ meson $\ket{B^{0}(\Delta t)}$ at the time of decay is given by (15), and the transition probability from the state $\ket{B^{0}(\Delta t)}$ to the state $\ket{f}$ is given by $|\braket{f|\hat{S}|B^{0}(\Delta t)}|^{2}$ using the unitary operator $\hat{S}$  corresponding to the S-matrix. Expanding the state $\ket{B^{0}(\Delta t)}$ in terms of the basis states $\ket{B^{0}}$ and $\ket{\bar{B}^{0}}$ as 
\begin{align}
    |B^{0}(\Delta t)\rangle=a(\Delta t)|B^{0}\rangle+b(\Delta t)|\bar{B}^{0}\rangle,
\label{testate}
\end{align}
we find that the probability becomes
\begin{align}
|\braket{f|\hat{S}|B^{0}(\Delta t)}|^{2}&=|a(\Delta t)\braket{f|\hat{S}|B^{0}}+b(\Delta t)\braket{f|\hat{S}|\bar{B}^{0}}|^{2},\nonumber\\
&=|a(\Delta t) A_{f}+b(\Delta t)\bar{A}_{f}|^{2},
\label{testate12}
\end{align}
where
\begin{align}
    A_{f}=\braket{f|\hat{S}|B^{0}},\quad \bar{A}_{f}=\braket{f|\hat{S}|\bar{B}^{0}},\label{def:dec_amplitude}
\end{align}
are the corresponding decay amplitudes from the basis states. Comparing~\eqref{testate12} with~\eqref{prob_bdecay}, we observe that if the condition
\begin{align}
    \frac{r^{\ast}}{s^{\ast}}=\frac{A_{f}}{\bar{A}_{f}},\label{rs_consistency_condition}
\end{align}
is satisfied, the probability~\eqref{testate12} becomes identical to the probability~\eqref{prob_bdecay} as a function of $r$ and $s$ up to normalization,
\begin{align}
|\braket{f|\hat{S}|B^0(\Delta t)}|^{2}\propto|\langle B_{\rm decay}|B^0(\Delta t)\rangle|^{2}.
\label{postselc}
\end{align}
This relation allows us to find the state $\bra{B_{\rm decay}}$ corresponding to the state $\bra{f}\hat{S}$ by applying the projection operator $\ket{B^{0}}\bra{B^{0}}+\ket{\bar{B}^{0}}\bra{\bar{B}^{0}}$ onto the space of $B$ meson as
\begin{align}
    \bra{B_{\rm decay}}&\propto \bra{f}\hat{S}(\ket{B^{0}}\bra{B^{0}}+\ket{\bar{B}^{0}}\bra{\bar{B}^{0}}),\nonumber\\
    &=A_{f}\bra{B^{0}}+\bar{A}_{f}\bra{\bar{B}^{0}},\nonumber\\
    &\propto r^{\ast}\bra{B^{0}}+s^{\ast}\bra{\bar{B}^{0}}.\label{postselected_B_state}
\end{align}
We thus see that choosing the decay mode $\ket{f}$ characterized by the parameters $r$ and $s$ via the amplitude $A_{f}$ and $\bar{A}_{f}$ amounts to choosing the state $\ket{B_{\rm decay}}$, \textit{i.e.}, the postselection. Our problem now boils down to finding how to choose the decay mode $\ket{f}$ with desired $r$ and $s$.
To demonstrate that this may in principle be possible, we recall that the Standard Model predicts that photons produced in the decay $b\to s\gamma$ are primarily left-handed~\cite{Grinstein_2005}.
For simplicity, in this paper, we assume that the involved photons are always left-handed, which implies that the photons produced from the decay of anti-$b$ quarks are always right-handed. It then follows that the combinations of the decayed particles are
\begin{align}
    |B^0\rangle \leftrightarrow |K^{\ast 0}\rangle|\gamma_{R}\rangle, \quad
    |\bar{B}^{0}\rangle \leftrightarrow |\bar{K}^{\ast 0}\rangle|\gamma_{L}\rangle,
\end{align}
where $|\gamma_{L}\rangle$ and $|\gamma_{R}\rangle$ represent the left-handed photon and the right-handed photon, respectively. 
Also, if we can assume, in effect, that this process preserves the $CP$ symmetry, $[\mathcal{CP}, \hat{S}] = 0$, then we have 
\begin{align}
    \hat{S}|B^0\rangle &= c\,|K^{\ast 0}\rangle|\gamma_{R}\rangle +\cdots,\label{u_def1}\\
     \hat{S}|\bar{B^0}\rangle &= c\,|\bar{K}^{\ast 0}\rangle|\gamma_{L}\rangle +\cdots,\label{u_def2}
\end{align}
with a common constant $c$, where $\cdots$ stands for other particles generated from the decay. 

This alludes us to consider the decay mode $\ket{f}$ by measuring the states   of the corresponding particles separately in the decay, \textit{i.e.}, we now have
\begin{align}
    \ket{f}=|K^{\ast}_{f}\rangle|\gamma_{f}\rangle, \label{final_state_explicitly}
\end{align} 
where $\ket{K^{\ast}_{f}}$ and $|\gamma_{f}\rangle$ are the states of the respective particles determined by the measurement.
Expanding $|K^{\ast}_{f}\rangle$ in terms of the basis states as
\begin{align}
    |K^{\ast}_{f}\rangle=\xi_{1}|K^{\ast 0}\rangle+\xi_{2}|\bar{K}^{\ast 0}\rangle,
     \label{postselect_kstar}
\end{align}
and similarly $|\gamma_{f}\rangle$ as
\begin{align}
    |\gamma_{f}\rangle=\eta_{1}|\gamma_{R}\rangle+\eta_{2}|\gamma_{L}\rangle,
    \label{postselect_photon}
\end{align}
and using~\eqref{u_def1},~\eqref{u_def2} together with the projection on the space of $B$ meson, we obtain
\begin{align}
    \left.\langle K^{\ast}_{f}|\langle \gamma_{f}|\hat{S}\,\right\rvert_{B\, {\rm meson}}
    &=\Bigl(\xi^{*}_{1}\eta^{*}_{1}\langle K^{\ast 0}|\langle \gamma_{R}|+\xi^{*}_{2}\eta^{*}_{1}\langle \bar{K}^{\ast 0}|\langle \gamma_{R}|\nonumber\\
    &\,\,+\xi^{*}_{1}\eta^{*}_{2}\langle K^{\ast 0}|\langle \gamma_{L}|+\xi^{*}_{2}\eta^{*}_{2}\langle \bar{K}^{\ast 0}|\langle \gamma_{L}|\Bigr)\hat{S}\nonumber \\
    &\quad(|B^0\rangle\langle B^0|+|\bar{B}^{0}\rangle\langle \bar{B}^0|) \nonumber\\
    &=c\, (\xi^{*}_{2}\eta^{*}_{2}\langle \bar{B}^{0}|+\xi^{*}_{1}\eta^{*}_{1}\langle B^{0}|).
    \label{postselect_gen}
\end{align}
From~\eqref{postselect_gen}, we find
\begin{align}
 A_{f}=\bra{K^{*}_{f}}\braket{\gamma_{f}|\hat{S}|B^{0}}&=c\, \xi^{*}_{1}\eta^{*}_{1},\\
    \bar{A}_{f}=\bra{K^{*}_{f}}\braket{\gamma_{f}|\hat{S}|\bar{B}^{0}}&=c\, \xi^{*}_{2}\eta^{*}_{2}.\label{absolute-r-s}
\end{align}
This implies that the consistency condition~\eqref{rs_consistency_condition} is satisfied if we choose the expanding parameters as
\begin{align}
    \frac{\xi_{1}\eta_{1}}{\xi_{2}\eta_{2}} = \frac{r}{s}.
    \label{par_con}
\end{align}
We thus see that, if we can freely select $\xi_{1}$, $\xi_{2}$, $\eta_{1}$, and $\eta_{2}$, for the decay mode $\ket{f}$ such that~\eqref{par_con} is fulfilled, we can freely select $r$ and $s$ for the postselected state $\ket{B_{\rm decay}}$, that is, the postselection can be made freely.

\begin{figure}[t]
    \includegraphics[width=8.5cm]{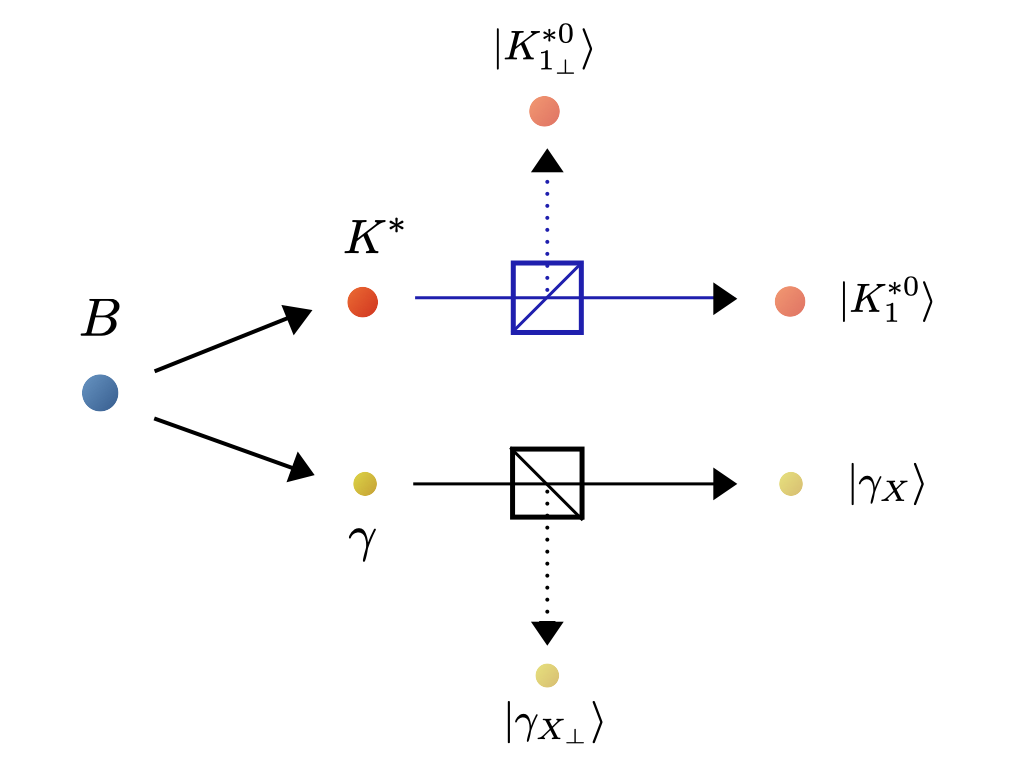}
    \caption{After the decay of the $B$ meson, the selection is performed for each of $K^{\ast}$ and $\gamma$.
    For the $K^{\ast}$, it is measured whether it is $|K^{\ast 0}_{1}\rangle$ given by~\eqref{KastbarKast} or $|K^{\ast 0}_{1_{\perp}}\rangle$, which is the orthogonal state. Similarly, for the  $\gamma$, we measure whether it is $|\gamma_{X}\rangle$ given by \eqref{eq:gamma_f} or its orthogonal state $|\gamma_{X_{\perp}}\rangle$. In the case of low-energy photons (visible light) often used in weak measurements, the beam splitter is commonly used for the postselection, while in the case of high-energy photons, a proper measurement of angular distribution of decayed particles may be considered alternatively.}
    \label{fig:postselection-scheme}
\end{figure} 

The parameters $\xi_{1}$, $\xi_{2}$, $\eta_{1}$, $\eta_{2}$ are not uniquely determined from the condition~\eqref{par_con}, and one possible choice is provided by choosing $|K^{\ast}_{f}\rangle = |K^{\ast 0}_{1}\rangle$, where 
\begin{align}
    |K^{\ast 0}_{1}\rangle=\frac{1}{\sqrt{2}}\left(|K^{\ast 0}\rangle+|\bar{K}^{\ast 0}\rangle\right).
    \label{KastbarKast}
\end{align}
This $CP$ eigenstate $|K^{\ast 0}_{1}\rangle$ is experimentally distinguished as it admits the decay mode into $K^{0}_{S}\pi^0$, which is not the case with its orthogonal state
\begin{align}
    |K^{\ast 0}_{1_{\perp}}\rangle=\frac{1}{\sqrt{2}}\left(|K^{\ast 0}\rangle-|\bar{K}^{\ast 0}\rangle\right).
\end{align}
Thus, the state $|K^{\ast 0}_{1}\rangle$ can be determined experimentally without ambiguity (with no component of $|K^{\ast 0}_{1_{\perp}}\rangle$), which implies that the postselection $|K^{\ast}_{f}\rangle = |K^{\ast 0}_{1}\rangle$ can be implemented  relatively easily.  With this choice, one easily finds from~\eqref{par_con} that the photon state~\eqref{postselect_photon} is now determined as $|\gamma_{f}\rangle =  |\gamma_{X}\rangle$ where
\begin{align}
   |\gamma_{X}\rangle = r|\gamma_{R}\rangle+s|\gamma_{L}\rangle.
    \label{eq:gamma_f}
\end{align}
Although it would be unrealistic to assume that the standard method of using the beam splitter for low-energy photonic systems is also available for high-energy photons, a more realistic method of choosing the polarized state~\eqref{eq:gamma_f} may be provided by measurement of angular distribution of decayed particles, which we shall mention in some detail in Section\,\ref{sec:Conclusion}. 
To sum up, we have learned that our desired postselection~\eqref{defbf} on the $B$ meson may be accomplished by choosing a corresponding decay mode appropriately (see Figure \ref{fig:postselection-scheme}).

Incidentally, we note that a similar argument applies to the $CP$ conjugate state,
\begin{align}
    |\bar{B}^0(\Delta t)\rangle=a(\Delta t)|\bar{B}^{0}\rangle+b(\Delta t)|{B}^{0}\rangle,
\end{align}
reaching essentially the same result~\eqref{postselc} with $|B^0(\Delta t)\rangle$ replaced by $|\bar{B}^0(\Delta t)\rangle$.

Note also that $K^{\ast}$ and $\gamma$ produced by the decay of the $B$ meson evolve over time after the production. While the state of the photon does not change as long as it flies in the vacuum, the $K^{\ast}$ state may change during the flight. Therefore, an additional operation is required to specify the state~\eqref{KastbarKast} taking into account this time evolution.

\section{Simulation Study}\label{sec:expSimStudy} 

\subsection{Introduction}\label{sec:expIntro} 

We shall now move on to discuss the weak measurement of the $B^{0} - \bar{B}^{0}$ mixing using the $B^{0}~\to~K^{*0}\gamma~\to~K^{0}_{S}\pi^{0}\gamma$ decay channel from the experimental point of view. 
Since our forgoing observation shows that the postselection after the $B^{0}~\to~K^{*0}\gamma~\to~K^{0}_{S}\pi^{0}\gamma$ decay can in effect be replaced by the postselection before the decay,  we focus on the latter in the following discussions.
Hereafter, we shall restrict ourselves to the type of flavor tagging carried out by identifying one of the species in the ($\mathcal{C} = -1$) entangled pair of $B$ mesons as adopted in the Belle~II experiment.  This requires us to take into account negative $\Delta t$ as well as positive ones, since the decay of one meson in the entangled pair may have already occurred when the tagging of the other meson is made. 
To accommodate this extension in the probability density function $P(\Delta t | B^{0} \to B_{\rm decay})$ in~\eqref{time-dist-wm}, we consider, in addition to the transition probability $|\langle B_{\rm decay}|B^{0}(\Delta t)\rangle|^{2}$ in~\eqref{prob_bdecay} for $\Delta t > 0$, the corresponding one $|\langle B^{0}|B_{\rm decay}(-\Delta t)\rangle|^{2}=|\langle B^{0}|e^{i\Delta t\hat{H}}|B_{\rm decay}\rangle|^{2}$ for $\Delta t<0$. This implies $\hat{H}\to \hat{H}^{\dag}$ when we go over from $\Delta t > 0$ to $\Delta t < 0$, and for our case where the Hamiltonian admits the form~\eqref{hami-b-meson}, this amounts to the parameter replacement $(p,q,\Gamma,\Delta\Gamma)\to(q^*,p^*,-\Gamma,-\Delta\Gamma)$.

In the probability density function $P(\Delta t | B^{0} \to B_{\rm decay})$ in~\eqref{time-dist-wm}, the effect of $CP$ violation in mixing arises from the difference between $p$ and $q$. In the SM, the difference of their absolute scalars $1 - |q|/|p|$ is expected to be small $\mathcal{O}(10^{-3})$~\cite{PDG_2020} in the $B^{0}-\bar{B}^{0}$ system.  Then, assuming $|q|/|p| = 1$ for simplicity, we find from
\eqref{time-dist-wm} and its $CP$ conjugate partner that the probability density functions reduce to 
\begin{widetext}
\begin{align}
P_{\rm phys}(\Delta t | B^{0} \to B_{\rm decay}) &=
\frac{e^{-\frac{|\Delta t|}{\tau}}}{2N} \left[ 1 + (2|r|^{2}-1)\cos{(\Delta m \Delta t) - 2|r|\sqrt{1 - |r|^{2}}}\sin{(\theta - \varphi)}\sin{(\Delta m \Delta t)} \right], \label{signalPdfB} \\ 
P_{\rm phys}(\Delta t | \bar{B}^{0}  \to \bar{B}_{\rm decay}) &=
\frac{e^{-\frac{|\Delta t|}{\tau}}}{2N} \left[ 1 + (2|r|^{2}-1)\cos{(\Delta m \Delta t) - 2|r|\sqrt{1 - |r|^{2}}}\sin{(\theta + \varphi)}\sin{(\Delta m \Delta t)} \right], \label{signalPdfBbar}
\end{align}
\end{widetext}
where $\tau = 1/\Gamma$ is the $B^{0}$ lifetime \cite{PDG_2020} and 
\begin{align}
N &= \tau \left( 1 + \frac{2|r|^{2} - 1}{1 + (\tau\Delta m )^{2}}  \right)
\label{signalPdfNorm}
\end{align}
is a normalization factor.  The formula~\eqref{signalPdfBbar} is obtained from~\eqref{signalPdfB} by putting $\varphi \rightarrow -\varphi$, which is a consequence of the interchange $p \leftrightarrow q$. The extension to $\Delta t < 0$ in this simplified case is achieved just by putting $\tau\to -\tau$ according to the note in the previous paragraph, resulting in the exponential factor in~\eqref{signalPdfB} and~\eqref{signalPdfBbar}.   
We note that in these formulae the only measurable parameter is $\varphi$ which corresponds to $2\phi_{1}$ in the SM (see Section~12 in~\cite{PDG_2020}), where $\phi_{1}$ is one of the interior angles of the unitarity triangle. 
If an extra phase arises from some new physics in $B\to K^{\ast} \gamma$ penguin diagram, the measured $\varphi$ will deviate from the $2\phi_{1}$.

\subsection{Belle II experiment}\label{sec:expBelle2} 

The Belle II experiment is currently operated at High Energy Accelerator Research Organization (KEK) in Japan, where one of the purposes is to measure the $CP$ violation in rare decay processes of $B$ mesons. The experiment will collect $550 \times 10^{8}$ $B\bar{B}$ pairs at 50~ab$^{-1}$, that are produced from resonance decays of $\Upsilon(4S)$ at the SuperKEKB. The SuperKEKB is an asymmetric energy electron-positron collider with 7~GeV and 4~GeV of electron and position beam energy, respectively. $\Upsilon(4S)$ is Lorentz-boosted with $\beta\gamma = 0.28$ along the direction of the electron beam. The Belle II detector consists of several sub-detectors surrounding the electron-positron collision point, which are designed to realize precise measurements of particle trajectories, production and decay vertices and particle identification.

The $B$ meson decaying to the $f_{\rm rec}$ final state, $K^{*0}\gamma$ in this study, is reconstructed from the decay particles detected by the Belle II detector.  The remaining particles in the detector, assumed as the particles from the other $B$ meson decaying to the $f_{\rm tag}$ final state, are used to tag the flavor of the $B$ meson decayed to the $f_{\rm rec}$ state, where the tagging time $t_{\rm tag}$ is the instant when the decay of the $B$ meson into $f_{\rm tag}$ occurred. The probability density functions of $B$ and $\bar{B}$ mesons, given by~\eqref{time-dist-wm} and its $CP$ conjugate partner, 
describe the time variation of the decays as a function of $\Delta t$ which can be determined from the displacement between $f_{\rm rec}$ and $f_{\rm tag}$ vertices along the beam axis, thanks to the asymmetric electron-positron beam energy.

\subsection{Event simulation}\label{sec:expEvtSimulation} 

Improvement of sensitivity to the $CP$ violation with weak measurement is evaluated in the $B^{0} \to K^{\ast0} \gamma$ decay as a benchmark by using pseudo-experiments. The $K^{*0} \to K^{0}_{S}\gamma$ decay mode is selected in addition to fulfill the consistency condition as discussed in Section~\ref{sec:Theoretical_model}. E. The number of events is assumed to be $550 \times 10^{8}$ $B\bar{B}$ in accordance with the expected integrated luminosity of the Belle~II experiment, where $B\bar{B}$ contains both $B^{0}\bar{B}^{0}$ and $B^{+}B^{-}$. The branching fractions of $\Upsilon(4S)~\to~B^{0}\bar{B}^{0}$ and $B^{0}~\to~K^{*0}\gamma$ decays are considered as 0.49 and $4.2~\times~10^{-5}$, respectively~\cite{PDG_2020}. Since the decay mode of $K_{S}^{0}$ is selected to be $K_{S}^{0} \to \pi^{+}\pi^{-}$ to reconstruct $B^{0}~\to~K^{*0}\gamma$ events via $K^{*0} \to K_{S}^{0}\pi^{0}$ decay in the Belle experiment, only the same decay mode is selected in this study. The branching fractions of $K^{*0} \to K_{S}^{0}\pi^{0}$ and $K_{S}^{0} \to \pi^{+}\pi^{-}$ are 0.17 and 0.69, respectively.

The efficiency to identify the flavor of $B$ mesons, $B^{0}$ or $\bar{B}^{0}$, decaying into $f_{\rm tag}$ is assumed to be 0.136, where the wrong tagging fraction is $0.02$~\cite{Kakuno:2004cf}. 
The reconstruction and selection efficiency will also be referred to the results in the Belle experiment, which is estimated to be 18.2\%~\cite{Horiguchi:2017ntw}. Finally, the signal yield is expected to be $3.2 \times 10^{3}$.

\begin{figure}[ht]
    \centering
    \includegraphics[width=8.5cm]{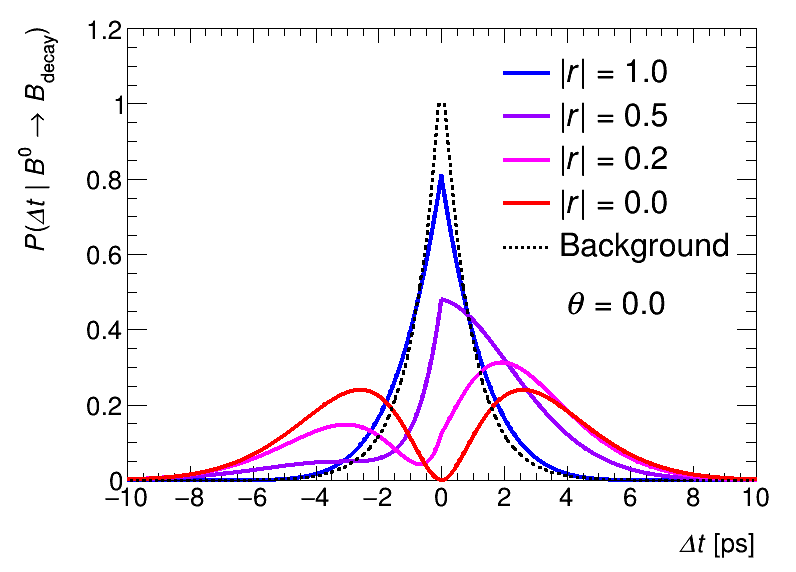}
    \caption{Time distribution of the signal and background, $P_{\rm phys}(\Delta t|B^{0}~\to~B_{\rm decay})$ and $P_{\rm bkg}(\Delta t)$, respectively. The solid lines are the signal distributions for various values of the postselection parameter, $|r|$, and the dashed line is the background.}
    \label{fig:pdf_B_decay_bg}
\end{figure}

Taking all these factors into account, we learn that the generation of simulated events follows the probability density function,
\fontsize{9.5pt}{0pt}\selectfont
\begin{widetext}
\begin{align}
\mathcal{P}(\Delta t|B^{0} \to B_{\rm decay}) = \int_{-\infty}^{\infty} d(\Delta t') \left[ f_{\rm phys} P_{\rm phys}(\Delta t'|B^{0} \to B_{\rm decay})R_{\rm phys}(\Delta t - \Delta t') + (1 - f_{\rm phys}) P_{\rm bkg}(\Delta t')R_{\rm bkg}(\Delta t - \Delta t') \right],
\label{dt_definition_B}
\end{align}
\end{widetext}
\normalsize
and its $CP$ conjugate partner $\mathcal{P}(\Delta t|\bar{B}^{0} \to \bar{B}_{\rm decay})$ in the corresponding form, both of which are characterized by 
the same fraction of the signal $f_{\rm phys} = 0.66$ taken from~\cite{Horiguchi:2017ntw}, together with $P_{\rm phys}$ and $P_{\rm bkg}$ which are 
the distributions of the signal equivalent to~\eqref{signalPdfB} and the background, respectively. 
The main background sources are the light quark pair production process ($e^{+}e^{-}~\to~q\bar{q}$ with $q~= u,d,s,c$) and the $e^{+}e^{-}~\to~B\bar{B}$ process where the $B$ meson decays into final states different from the signal leading to misidentification. The profile of the background distribution is empirically determined from the results in the Belle experiment as 
$P_{\rm bkg}(\Delta t') ~=~\frac{1}{\tau_{\rm bkg}}e^{-\left| \Delta t' \right|/\tau_{\rm bkg}}$,
where $\tau_{\rm bkg}$ is set to be 0.896 ps. Figure~\ref{fig:pdf_B_decay_bg} shows the time distribution of the signal, $P_{\rm phys}(\Delta t|B^{0}~\to~B_{\rm decay})$ and backgrounds, $P_{\rm bkg}(\Delta t)$. The postselection can be a powerful tool to discriminate the signal from backgrounds because the separation of them is more significant as $|r|$ goes to zero.

\begin{figure}[htbp]
\begin{center}
\includegraphics[width=8.5cm]{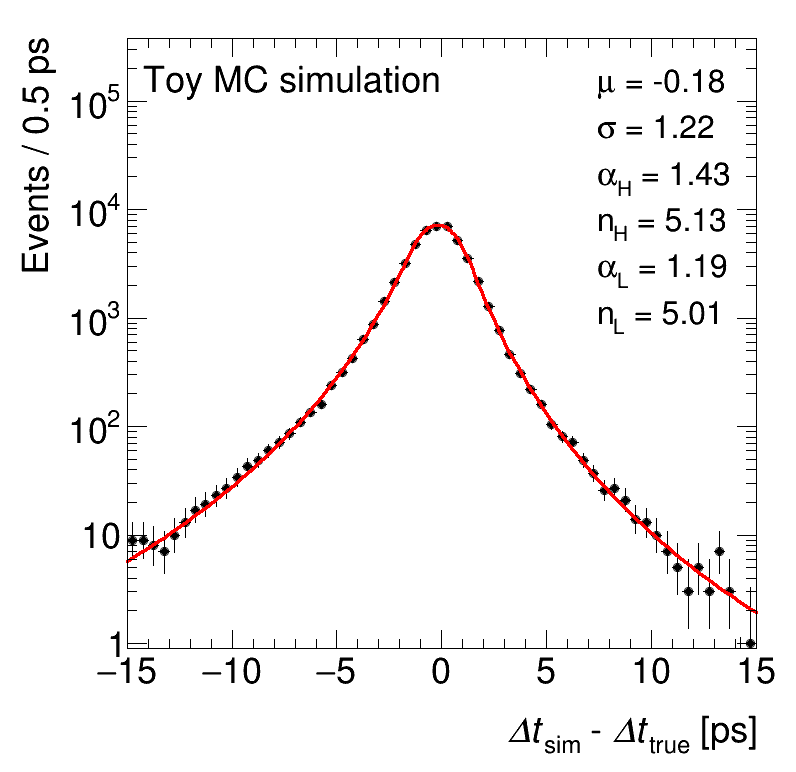}
\caption{Distribution of the $(\Delta t_{\rm sim}$ and $\Delta t_{\rm true})$ differences (residuals).
The points are simulated.
The red line is the fitted function which is empirically modeled by the double-sided crystal ball function with parameters shown above.}
\label{fig:resoFunction}
\end{center}
\end{figure}

The time resolution functions for the signal $R_{\rm phys}$ and background $R_{\rm bkg}$ are evaluated separately and convoluted to each distribution, while the same profile is used both for $R_{\rm phys}$ and $R_{\rm bkg}$. The time resolution depends on the position resolution of decay vertices which should be considered to reproduce the timing response in the actual experiment. Figure~\ref{fig:resoFunction} shows the residual of $\Delta t_{\rm sim}$ and $\Delta t_{\rm true}$, where $\Delta t_{\rm sim}$ is the simulated $\Delta t$ where the position resolution of decay vertices is taken into account, based on the Belle experiment~\cite{Horiguchi:2017ntw}, and $\Delta t_{\rm true}$ is the true value of $\Delta t$ without the detector response. 

To obtain the resolution function of $\Delta t$, Figure~\ref{fig:resoFunction} is fitted by an empirical function, called \lq double-sided crystal ball function\rq. The double-sided crystal ball function consists of a combination of a Gaussian and power law tails, defined as

\begin{widetext}
\begin{align}
R(x) & = N \times 
	\begin{cases}
		e^{-\frac{X^{2}}{2}}, & \hbox{for} \quad -\alpha_{L} \le X \le \alpha_{H}, \\
		e^{-\frac{\alpha_{L}^{2}}{2}}\left[\frac{\alpha_{L}}{n_{L}}(\frac{n_{L}}{\alpha_{L}} - \alpha_{L} - X)\right]^{-n_{L}}, & \hbox{for} \quad X < -\alpha_{L}, \\
		e^{-\frac{\alpha_{H}^{2}}{2}}\left[\frac{\alpha_{H}}{n_{H}}(\frac{n_{H}}{\alpha_{H}} - \alpha_{H} + X)\right]^{-n_{H}}, & \hbox{for} \quad X > \alpha_{H},
	\end{cases} 
\end{align}
\end{widetext}
with the shorthand,
\begin{align}
	X  = \frac{x - \mu}{\sigma} \label{resoFunction},
\end{align}
where $N$ is the normalization factor, $\mu$ and $\sigma$ are the mean and width of the Gaussian, $\alpha_{H}$ and $\alpha_{L}$ are the positions of transition from the Gaussian to the power low tails on the higher or lower sides, and $n_{H}$ and $n_{L}$ are the exponents of the high and low tails, respectively. 

The resolution functions are well modeled by fitting the simulated data under the parameters shown there.  

The postselection parameters $|r|$ and $\theta$ in~\eqref{signalPdfB} and~\eqref{signalPdfBbar} represent the mixing ratio and the phase difference between $|B^{0}\rangle$ and $|\bar{B}^{0}\rangle$ states, respectively. They are the parameters of the postselection that have to be determined by event selections in the experiment. Although their determination is an important technical subject in the actual experiment, for simplicity we assume to have some means to determine them as we wish. The selection efficiencies of $|r|$ and $\theta$ are both assumed as 100\% in this study, so that dependence of sensitivity to $\varphi$ only on these parameters can be investigated. The method how to determine these postselection parameters is discussed in Section~\ref{sec:Conclusion}.

\subsection{Signal extraction}\label{sec:expSigExtract} 

In order to evaluate the measurement accuracy of the $CP$ parameter $\varphi$, an unbinned maximum likelihood fit is performed for the distributions obtained in the pseudo-experiments. The probability density function given in~\eqref{dt_definition_B} depends on the fitting parameter $\varphi$ as well as systematic uncertainties coming from the three sources:
the background estimation, the timing scale and the timing resolution, which correspond to 
$f_{\rm phys}$, $\mu$ and $\sigma$ appeared in~\eqref{dt_definition_B} with~\eqref{resoFunction}.
A set of systematic uncertainties is parameterized by the so-called nuisance parameters $\theta^{\rm nuis}$ that follow the standard Gaussian distribution $G(\theta^{\rm nuis})$ (see Section~40 in~\cite{PDG_2020}).
Then, the likelihood function $L$ is defined by the product,
\begin{align}
L = \mathcal{P}(\Delta t, \theta_{f_{\rm phys}}^{\rm nuis}, \theta_{\mu}^{\rm nuis}, \theta_{\sigma}^{\rm nuis})\, G(\theta_{f_{\rm phys}}^{\rm nuis})\, G(\theta_{\mu}^{\rm nuis})\, G(\theta_{\sigma}^{\rm nuis}),
\end{align}
where these $G(\theta^{\rm nuis})$ provide constraints on the systematic uncertainties~\cite{Cowan:2010js}.

The unbinned maximum likelihood fit is performed simultaneously for the distributions of $B^{0}\to B_{\rm decay}$ and $\bar{B}^{0}\to \bar{B}_{\rm decay}$. Systematic uncertainties estimated in the Belle experiment~\cite{Horiguchi:2017ntw,Ushiroda:2006fi} are used in this study. The dominant systematic uncertainties derive from the background estimation, the timing scale and the resolution of uncertainties in the position of the reconstructed vertices. Only these uncertainties are considered in the unbinned maximum likelihood fit.

\subsection{Analysis result}\label{sec:expResult} 

A pseudo-experiment is carried out and repeated a thousand times to estimate the measurement precision of $\varphi$ per particular set of the post selection parameters $|r|$ and $\theta$, in which events are randomly generated, following the probability density functions,~\eqref{dt_definition_B} and its $CP$ conjugate partner. These parameters are scanned in the range of $0.1 \le |r| \le 0.9$ with the intervals of 0.1 and similarly in the range of  $-180 \le \theta \le 180$ degrees with the intervals of 36 degrees. The initial value of $\varphi$ is set to 44.4 degrees which corresponds to $\varphi = 2\phi_{1}$ with the world average value of $\phi_{1} = 22.2 \pm 0.7$ degrees~\cite{Amhis:2019ckw}. The unbinned maximum likelihood fit is performed to each pseudo-experiment. 

\begin{figure}[htbp]
\begin{center}
\includegraphics[width=8.5cm]{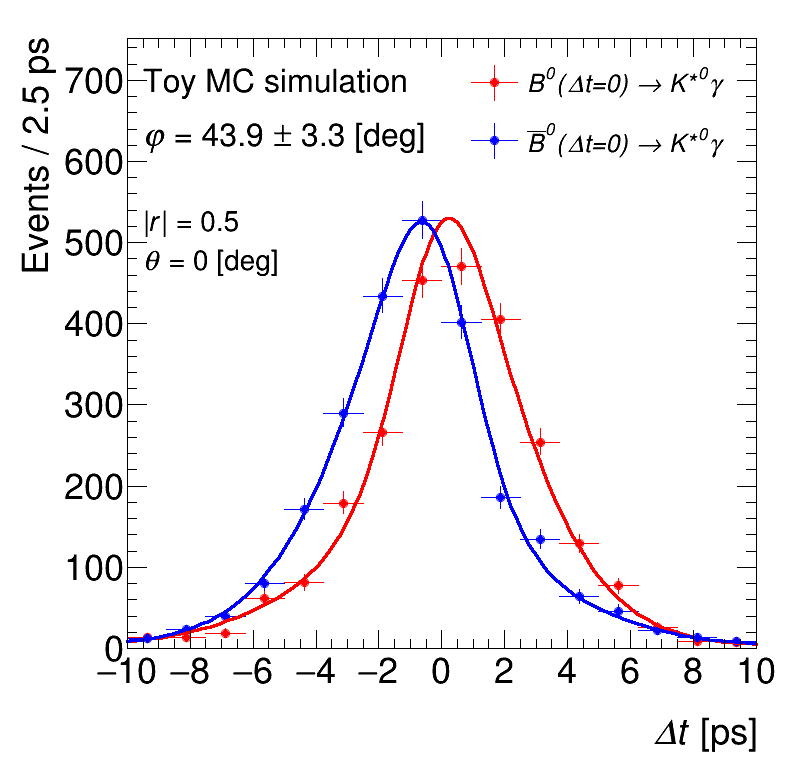}
\caption{Distribution of $\Delta t$ for $B^{0}$ and $\bar{B}^{0}$ initial states.
The postselection parameters $|r|$ and $\theta$ are selected to be $0.5$ and $0.0$ degrees, respectively, which minimize the total uncertainty of measured $\varphi$. The red and blue lines represent the best fit probability density functions, $\mathcal{P}(\Delta t~|~B^{0}~\to~B_{\rm decay})$ and $\mathcal{P}(\Delta t~|~\bar{B}^{0}~\to~\bar{B}_{\rm decay})$, respectively.}
\label{fig:fitCategory}
\end{center}

\end{figure}

Figure~\ref{fig:fitCategory} shows the $\Delta t$ distributions when $|r|$ and $\theta$ are selected to be $0.5$ and $0.0$ degree, respectively, which minimize the total uncertainty of measured $\varphi$. The fitted functions are shown in the plot, which are normalized to the number of events in each distribution. The shift of the peak position between $B^{0} \to K^{\ast0} \gamma$ and $\bar{B}^{0} \to K^{\ast0} \gamma$ shows the effect of the $CP$ violation in the $B^{0} - \bar{B}^{0}$ mixing with the $CP$ parameter $\varphi$.

\begin{figure}[htbp]
\begin{center}
\includegraphics[width=8.5cm]{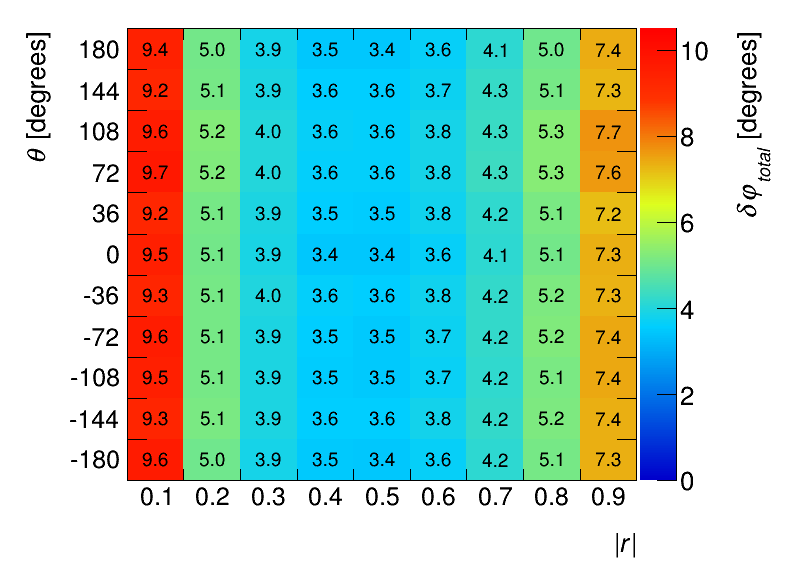}
\caption{Total uncertainty of the $\varphi$ measurement ($\delta \varphi_{\rm total}$) as a function of $|r|$ and $\theta$ used as the input values.}
\label{fig:errorScan}
\end{center}
\end{figure}

Figure~\ref{fig:errorScan} shows the total uncertainty of measured $\varphi$ as a function of $|r|$ and $\theta$ which are used as the input values. This plot indicates that the measurement precision is improved by 20\% with $|r| = 0.5$ compared to the case in which the final state is the $CP$ eigenstate with $|r| = \frac{1}{\sqrt{2}} \simeq 0.7$.  Note that the precision is virtually insensitive to $\theta$. The systematic uncertainty is in the range of 0.2 - 1.1 degrees while the statistical uncertainty varies from 3.4 to 9.7 degrees, depending on the postselection parameters, indicating the dominance of the statistical uncertainty in our analysis.

The effective lifetime can be prolonged by selecting small $|r|$ in the postselection as shown in Figure~\ref{fig:weak_value_amplification_of_lifetime}. The level of contamination of background peaked at $\Delta t = 0$ tends to diminish as the effective lifetime becomes longer, {\it i.e.}, the uncertainty on background estimation can be reduced for smaller $|r|$ up to $\sim 0.5$. The uncertainty of $\varphi$, however, starts to grow for $|r|$ smaller than $\sim 0.5$, because the term containing $\sin(\theta - \varphi)$ in~\eqref{signalPdfB} and~\eqref{signalPdfBbar} becomes negligible then. This effect can be seen in Figure~\ref{fig:errorScan}. 
The fitting method is validated with the number of pseudo-experiments we have performed, in which the results are consistent with the input value within the statistical uncertainty.

\section{Conclusion and Discussions}\label{sec:Conclusion}

\begin{figure}
    \centering
    \includegraphics[width=8.5cm]{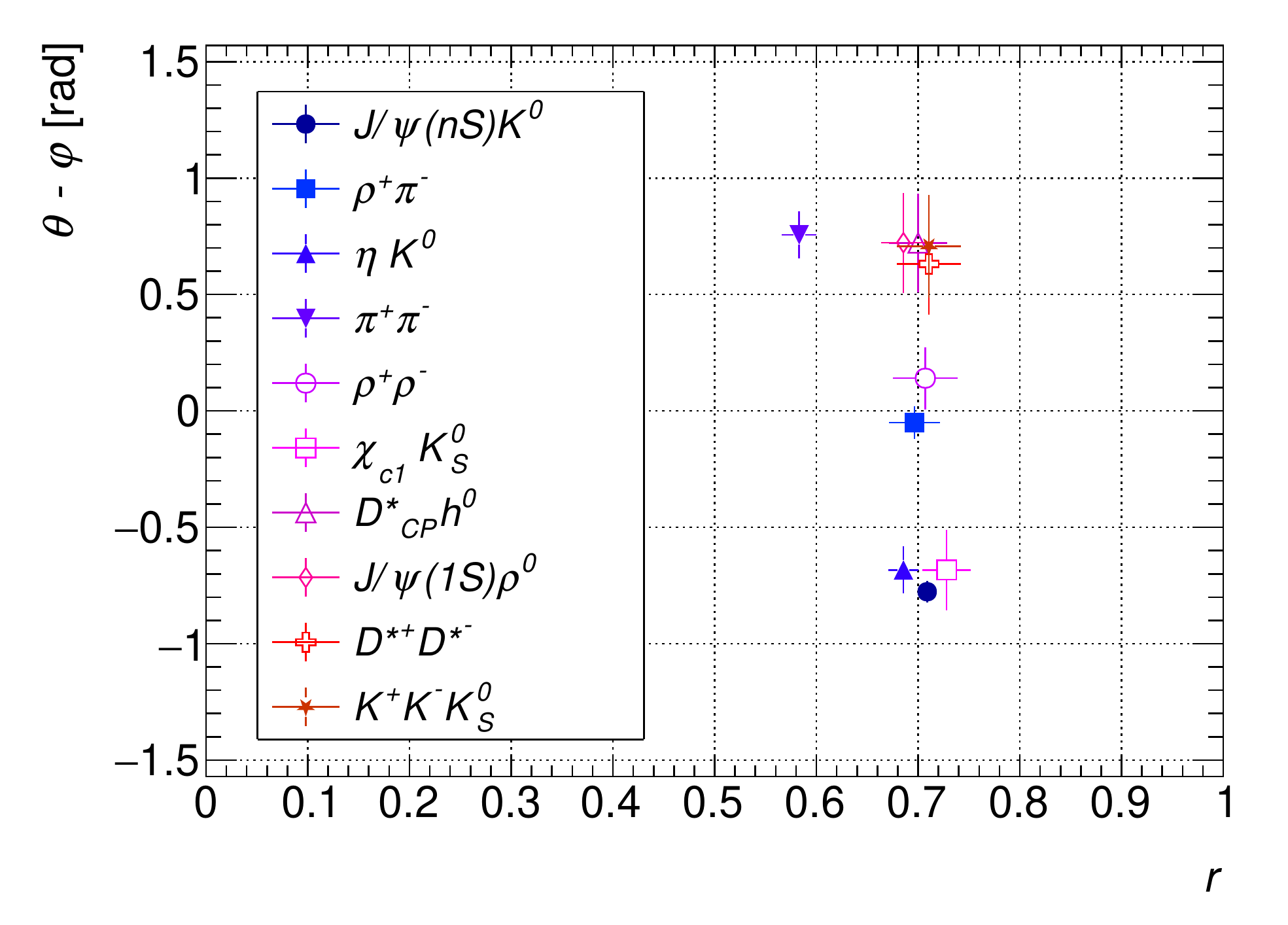}
    \caption{Distribution of 10 major postselections performed in the previous measurements of $CP$ violation in $B$ decays. 
The values are estimated from the data in PDG \cite{PDG_2020}, pp.63-64, under the identification $C=2|r|^2-1$, $S=-2|r|\sqrt{1-|r|^2}\sin(\theta-\varphi)$ obtained from our formulae \eqref{signalPdfB}, \eqref{signalPdfBbar} in Section \ref{sec:expSimStudy}, for parameters $r,\theta$ of postselections.  The range in the choices of $|r|$ is seen to be restricted to a narrow band around $|r| = 0.7 (\approx 1/\sqrt{2})$. For a fixed value of $\varphi$, the distribution along the vertical axis (plotted in the unit of radian) amounts to the distribution of $\theta$.}
    \label{r_theta_dat}
\end{figure}

We have shown that the effect of postselection is clearly seen in the time distribution of the $B$ meson decay, and that the postselection has the effect of amplifying both the effective lifetime and the $CP$ violation. In our simulation study, which assumed that the measurement is performed under the parameters of the Belle II experiment, the effective lifetime is found to be prolonged by $2.6$ times, and the accuracy of $CP$-violating parameters can also be improved when it is realized in the decay mode we considered.  
Although these results may not be as great as one might wish to see, we believe that our case study presented in this paper certainly indicates the potential use of weak measurement for high energy physics, especially with respect to the exploitation of the freedom of postselection.   
Our attempt provides a first systematic study for such exploitation, given that the freedom of postselection has not been utilized comprehensively for amplifying the prospected signals before.  This can be seen clearly in Figure \ref{r_theta_dat}, where postselections realized by previous measurements form only a limited class in the possible range leaving a vast room of freedom unexplored.

Our case study may lead to more interesting results in view of the new physics related to the decay mode $B \to K^{*} \gamma$. If the effect signalling the new physics associated with the right-handed current is sizable to $B\to K^{*} \gamma$, we should be able to measure the finite effect in the coefficient of $\sin(\Delta m \Delta t)$. However, with the conventional method, we cannot separate the absolute amplitude and the phase in the new physics.  The separation may become possible with the weak measurement if it is performed under different sets of parameters $(r, \theta)$, elucidating the details of the new physics we have found.

\begin{figure*}[htbp]
\begin{center}
\begin{tabular}{c}
\begin{minipage}{0.5\hsize}
\begin{center}
\includegraphics[width=8.5cm]{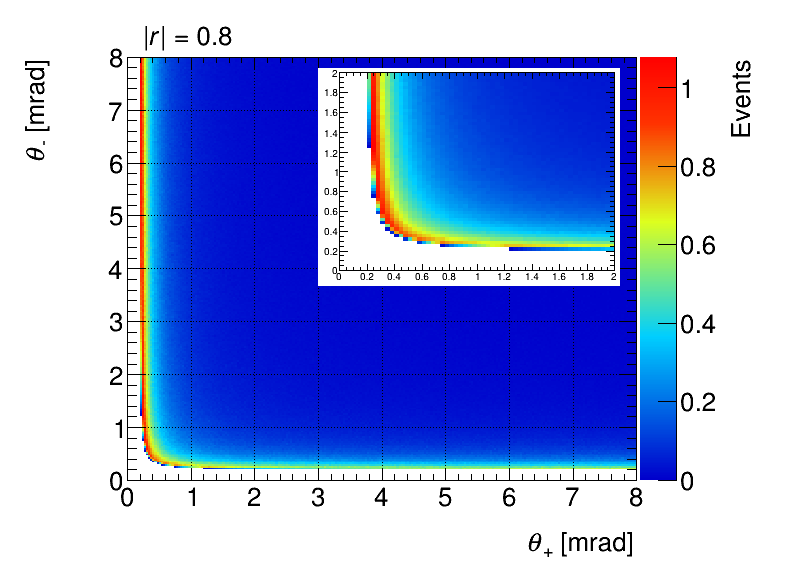}
\end{center}
\end{minipage}
\begin{minipage}{0.5\hsize}
\begin{center}
\includegraphics[width=8.5cm]{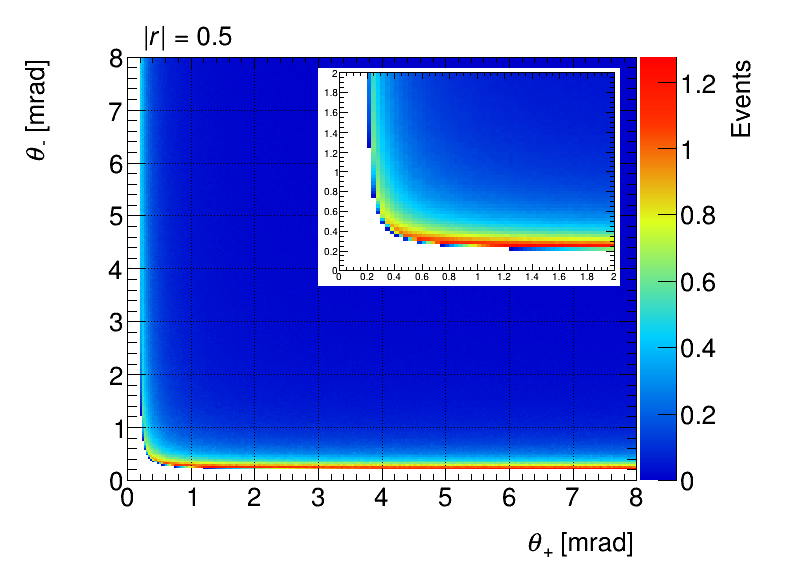}
\end{center}
\end{minipage}
\end{tabular}
\caption{Distribution of the polar angle of charged leptons from the photon conversion for $|r|$ = 0.8 (left) and 0.5 (right). The $z$-axis is normalized to the number of the expected events in the Belle II experiment. The inset shows the same distribution zooming up the range of $\theta_{\pm}$.}
\label{fig:angularDist}
\end{center}
\end{figure*}

In high energy physics, due to the short coherence length compelled by the large momentum involved in the experiment, one normally supposes that approaches which exploit 
superpositions of states such as the weak value amplification will not be available.  However, our analysis shows that the coherence time, rather than the coherence length, is a deciding factor for the amplification to work, as we have demonstrated in our analysis for the $B^0 - \bar{B}^0$ system. Similar positive outcomes will be expected generally for systems which admit transitions characterized by the average decay width $\Gamma$ and the difference of the energy $\Delta m$. 

The success of the weak value amplification hinges on the capability of exploiting the freedom of postselection. In high energy experiments, this must be realized at the level of final decay modes which are actually measured.  
In our case, this is achieved by considering measurements on the particles $K^{*}$ and $\gamma$ generated from the $B$ decay, in which our proper choice of the final states satisfies the consistency condition to reproduce the transition probability evaluated directly with the postselected states.  As such, the amount of data to be used for amplification may be considerably reduced rendering the statistical accuracy poorer.  
It is thus desirable to look for other decay modes, fulfilling the same consistency condition to improve the statistics and eventually the accuracy as well.  We also remark that, in order to improve the sensitivity of our weak measurement further, those contributions from the direct and indirect $CP$ violation, which we have neglected for simplicity in our analysis, should also be taken into account.

\begin{figure*}[htbp]
\begin{center}
\begin{tabular}{c}
\begin{minipage}{0.5\hsize}
\begin{center}
\includegraphics[width=8.5cm]{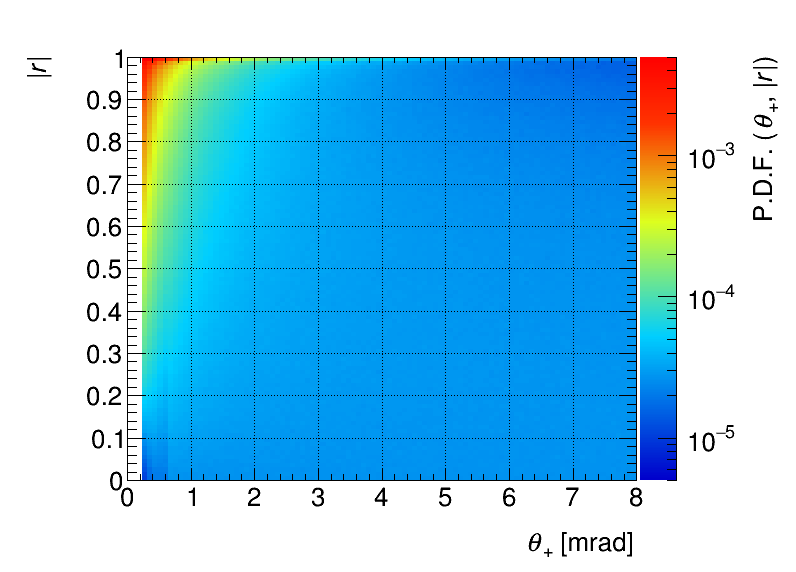}
\end{center}
\end{minipage}
\begin{minipage}{0.5\hsize}
\begin{center}
\includegraphics[width=8.5cm]{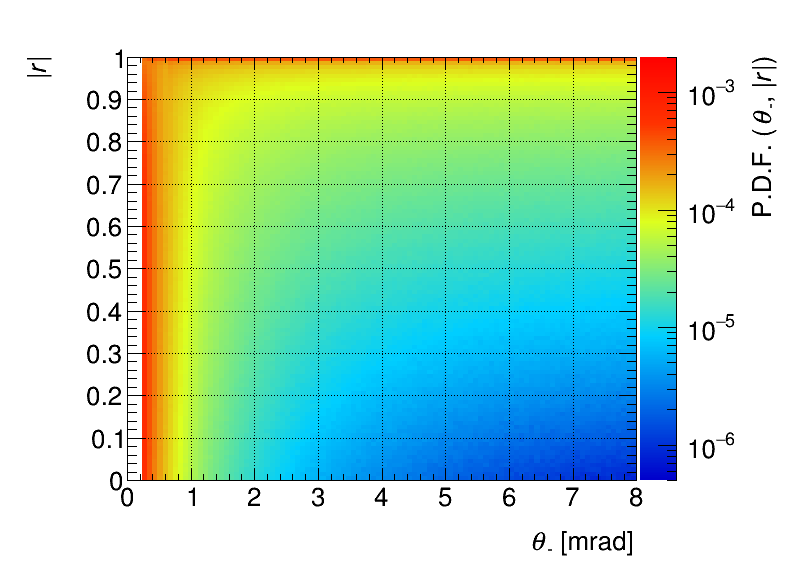}
\end{center}
\end{minipage}
\end{tabular}
\caption{Probability density function of $|r|$ and the polar angle of charged leptons from the photon conversion: $\theta_{+}$ (left) and $\theta_{-}$ (right).}
\label{fig:angularRratioDist}
\end{center}
\end{figure*}

For actual experimental implementation, one has to devise a workable procedure to achieve the postselection with a given set of parameters, $\vert r\vert$ and $\theta$, specifying the superposition of $|B^{0}\rangle$ and $|\bar{B}^{0}\rangle$.  
The circular polarization of the photon is considered as a key element for the postselection for $B \to K^{\ast} \gamma$ in our study. Identification of the photon polarization may be realized by utilizing the fact that angular distribution of charged lepton pairs from the photon conversion depends on it ~\cite{Kolbenstvedt1964CircularPP,PhysRev.114.887}.
This may allow us to select one of the postselection parameter $|r|$, although the probability of photon conversion is typically low: approximately $6\%$ at Belle II.
In addition to photon conversion events, contribution from $B^{0} \to K^{*}l^{+}l^{-}$ channel with $(9.9^{+1.2}_{-1.1}) \times 10^{-7}$ of the measured branching ratio~\cite{PDG_2020} can be added. 
Figure~\ref{fig:angularDist} shows the distributions of the polar angle $\theta_{\pm}$ of the positive and negative charged leptons created from the photon conversion with respect to photon momentum in the lab frame for $|r| = 0.8$ (left) and $0.5$ (right), assuming that photon energy is $2.6~\mathrm{GeV}$. The fact that the approximation $E_{\pm}\theta_{\pm}/m_{l} \sim 1$ maximizes sharply the amplitude of photon conversion~\cite{Bishara:2015yta} is used in this calculation, where $E_{\pm}$ is energy of the positive and negative charged leptons, and $m_{l}$ is their mass.
Figure~\ref{fig:angularRratioDist} shows the probability density functions of $|r|$ and $\theta_{\pm}$.
The dependence of angular distribution on $|r|$ will allow us to implement the postselection for $|r|$ by selecting a certain region of the distribution. 
For example, from the events distributed at low $\theta_{+}$ and $\theta_{-}$ for small and large $|r|$, respectively, the efficiency is estimated as $\sim$45$\%$ by selecting $\theta_{-} < 2$ mrad.
Once $|r|$ is identified, the procedure described in~\cite{Bishara:2015yta} then furnishes a possible means to determine the other postselection parameter $\theta$. Although the principle to determine the photon circular polarization has been discussed already long time ago, its experimental demonstration has not been done so far. The Belle II experiment has a good transverse momentum ($p_{\mathrm{T}}$) resolution of $0.3$-$0.6\%$ for charged particles with $p_{\mathrm{T}} = 0.5$-$1.0~\mathrm{GeV}/c$~\cite{Adachi:2008da}, which will result in the angular resolution better than $1.2\%$.
This may be enough to identify the photon circular polarization with the method discussed above. 


Furthermore, the technique of the developing machine learning may also be useful to find the optimal $\vert r\vert$ and $\theta$ and thereby ascertain the benefit of the weak measurement systematically.

Finally, we wish to argue that, in order to explore a new horizon beyond the conventional means of experimentation of high energy physics, it is important to actively utilize the properties of elementary particles as genuine quantum states.  We hope that the present study encourages those who share the goal in the same direction.

\bigskip
\begin{acknowledgments}
We would like to thank Takanori Hara, Yutaka Shikano, Masataka Iinuma, Satoshi Iso, Kazutaka Sumisawa and Katsuo Tokushuku for useful discussions. This work was supported by JSPS KAKENHI Grant Number 20H01906. 
\end{acknowledgments}

\bibliography{main}
\end{document}